\documentclass[11pt]{article}

\usepackage{amsfonts}
\usepackage{amsmath}
\usepackage{amssymb}
\usepackage{amsthm}
\usepackage{xcolor}
\usepackage[T1]{fontenc}
\usepackage[scaled]{helvet}
\usepackage{rotating}
\usepackage{multirow}
\usepackage{natbib}
\usepackage{appendix}
\usepackage{enumitem}
\usepackage{epstopdf}
\usepackage{bm}
\usepackage{slashbox}
\usepackage[colorlinks=true,allcolors=blue]{hyperref}
\usepackage{multirow}
\usepackage{float}
\usepackage{multicol}
\usepackage{arydshln}
\usepackage{bigints}
\usepackage{subfig}
\usepackage{gensymb}
\usepackage{bbm}
\usepackage{enumitem}
\usepackage{arydshln}

\DeclareMathAlphabet{\mathdutchcal}{U}{dutchcal}{m}{n}

\makeatletter
\@addtoreset{equation}{section}

\makeatother

\newtheorem{theorem}{Theorem}
\newtheorem{remark}{Remark}
\newtheorem{corollary}{Corollary}
\newtheorem{algorithm}{Algorithm}

\DeclareMathAlphabet{\mathpzc}{OT1}{pzc}{m}{it}

\newenvironment{steps}
 {\begin{enumerate}[label=Step \arabic*:,leftmargin=*,align=left]}
 {\end{enumerate}}
\newenvironment{substeps}
 {\begin{enumerate}[label=Step \arabic{enumi}.\alph*:,leftmargin=*,align=left]}
 {\end{enumerate}}

\begin{document}

\title{A reliable data-based smoothing parameter selection method for circular kernel estimation}

\author{Jose Ameijeiras-Alonso$^{1}$
}
\date{%
		CITMAga, Department of Statistics, Mathematical Analysis and Optimization, Universidade de Santiago de Compostela, Spain.
}
\footnotetext[1]{Supported by Grant PID2020-116587GB-I00 funded by MCIN/AEI/10.13039/501100011033 and the Competitive Reference Groups 2021-2024 (ED431C 2021/24) from the Xunta de Galicia. The author is most grateful to Rosa M. Crujeiras for helpful suggestions and comments on an earlier draft.}
\maketitle

\begin{abstract}
A new data-based smoothing parameter for circular kernel density (and its derivatives) estimation is proposed. Following the plug-in ideas, unknown quantities on an optimal smoothing parameter are replaced by suitable estimates. This paper provides a circular version of the well-known Sheather and Jones bandwidths (DOI: 10.1111/j.2517-6161.1991.tb01857.x), with direct and solve-the-equation plug-in rules. Theoretical support for our developments, related to the asymptotic mean squared error of the estimator of the density, its derivatives, and its functionals, for circular data, are provided. The proposed selectors are compared with previous data-based smoothing parameters for circular kernel density estimation. This paper also contributes to the study of the optimal kernel for circular data. An illustration of the proposed plug-in rules is also shown using real data on the time of car accidents.
\end{abstract}

\noindent%

\section{Introduction}

Circular data are observations that can be represented on the unit circumference and where periodicity must be taken into account. Classic examples appear when the goal is to model orientations or a periodic phenomenon with a known period. Several applications of circular data can be found, e.g., in \cite{Ley2018} or \cite{sengupta2022}. The complicated features that circular data can exhibit on real data applications lead to several new flexible models in the statistical literature. A recent review of flexible parametric circular distributions can be found in \cite{ameijeiras2022}. 

When trying to obtain a flexible fit of the density function, an alternative to parametric models is the kernel density estimation. Kernel density estimation for circular data dates back to \cite{beran1979} and \cite{Hall1987}, while the estimator of the density derivatives was studied by \cite{Klemela2000} and \cite{DiMarzio2011}. It is well known that the choice of the smoothing parameter is critical, when using these kernel methods.

In the usual linear inferential framework, where random variables are supported on the Euclidean space, one can find many approaches for selecting the ``best'' data-driven bandwidth parameter \citep[see, e.g.,][for a discussion on this topic]{Jones1996}. Due to its good performance, one of the most-employed bandwidth selectors is the plug-in bandwidth proposed by \cite{sheather1991}. The relevance of this plug-in selector is evident from the impressive number of citations that \cite{sheather1991} have received, and although other authors have introduced new bandwidth selectors, none of the proposals outperforms, in general, their plug-in selector.

There exists some literature on smoothing parameter selection for circular data, some of these ideas are based on cross-validation \citep{Hall1987}, rule-of-thumb \citep{Taylor2008}, adaptive-mixture of von Mises \citep{Oliveira2012}, or bootstrap \citep{DiMarzio2011} techniques. But none of the approaches introduced so far presents an outstanding performance with respect to its competitors, as is the case with the proposal by \cite{sheather1991} in the linear case. Hence, the goal of this paper is to provide the needed theory to derive an algorithm that replicates the idea of the two-stage direct and solve-the-equation plug-in bandwidth selectors for circular kernel density estimation. In addition, the developed theory can be also employed when estimating the density derivatives. 

Regarding the kernel choice, most of the results for circular density estimation fix the kernel to be the von Mises density function. In this paper, we study the asymptotic results for a more general class of kernels. The developed results allow us to obtain the optimal kernels in the circular estimation context.

This paper is organized as follows. Section~\ref{circular_kdde} is devoted to the definition of the circular kernel density derivative estimate. Also, the key function needed to derive the optimal smoothing parameter is introduced in this section. For a general circular kernel, the asymptotic mean integrated squared error and the optimal smoothing parameter of the derivative estimator are derived in Section~\ref{asymptotic_kdde}. Section~\ref{circular_functionals} is devoted to the estimation of density functionals, needed to derive the plug-in rules. In Section~\ref{kernel_choice}, we discuss the kernel choice. Section~\ref{implementation_practice} provides all the needed details to compute the two-stage direct and solve-the-equation plug-in smoothing selectors. A simulation study showing that the proposed rules provide a competitive smoothing parameter is given in Section~\ref{simulation_study}. Section~\ref{real_data} uses a real data example to show the applicability of the proposed selector. Some final remarks are provided in Section~\ref{concluding}. Section~\ref{software} describes the software that implements the proposed smoothing parameters. The proofs of the theoretical results appear in Appendices~\ref{proofs_ker_der} and~\ref{proofs_functional}. 

\section{Circular kernel density derivative estimation}\label{circular_kdde}

Given a circular random sample in angles $\Theta_1, \ldots, \Theta_n \in [-\pi,\pi)$, with associated density function $f$, the circular kernel density estimation \citep[see, for example,][]{Oliveira2012} can be defined as follows
\begin{equation}\label{kernel}
\hat{f}_\nu(\theta)=\frac{1}{n} \overset {n} {\underset {i=1} \sum} K_\nu\left(\theta-\Theta_i\right),
\end{equation}
where $K_\nu$ is the circular kernel with smoothing parameter $\nu \in [0,1]$, denoting by $\nu$ the mean resultant length (differently from the previous literature, where $\nu$ usually stands for a concentration parameter). This allows us to establish a theory that will be valid for both, kernels depending on the concentration, such as the von Mises, or on the mean resultant length, such as the wrapped normal (see Section~\ref{kernel_choice} for a formal introduction of these kernels).

The estimator \eqref{kernel} can be easily extended, when the objective is estimating the $r$--th derivative of $f$, denoted by $f^{(r)}$. In that case, the estimator of $f^{(r)}$ can be defined as \citep{DiMarzio2011},

\begin{equation}\label{kernel_der}
\hat{f}^{(r)}_\nu(\theta)=\frac{1}{n} \overset {n} {\underset {i=1} \sum} K^{(r)}_\nu\left(\theta-\Theta_i\right).
\end{equation}

The first question is which function can be employed as a kernel? In this case, we will assume that $K$ satisfies the standard conditions that one can find for the canonical linear kernel \citep[][Ch. 2]{WandJones1995}. In particular, $K$ will be a circular density, (reflectively) symmetric about zero, unimodal and square-integrable in $[-\pi,\pi)$. Note that these last conditions ensure that the kernel $K$ has the following convergent Fourier series representation \citep[see][Section 4.2]{Mardia1999a},

\begin{equation}\label{convergent_Fourier}
K_\nu (\theta) = \frac{1}{2\pi} \left( 1+ 2\sum_{j=1}^{\infty} \alpha_{K,j}(\nu) \cos (j\theta) \right),
\end{equation}
where only the values of $\alpha_{K,j}(\nu)\in [0,1]$, for $j\in\{1,2,\ldots\}$, depend on the employed kernel and the smoothing parameter $\nu$.

Secondly, a crucial element in kernel density estimation is the smoothing parameter. Generally, this parameter is taken as a non-random sequence, depending on the sample size, and then as a fixed value for a sample realization. In this work, we will introduce the function $h$ depending on $\nu$,

\begin{equation}\label{h_value}
h \equiv h_{K}(\nu)= \frac{\pi^2}{3}+4 \sum_{j=1}^{\infty} \frac{(-1)^j \alpha_{K,j}(\nu)}{j^2}.
\end{equation}

Then, $h$ can be seen as the \textit{bandwidth} and the sequence of numbers $h_{n} \equiv h_{K}(\nu_n)$ must satisfy the following condition $\lim_{n\rightarrow\infty} h_{n} =0$.

\section{Asymptotic results}\label{asymptotic_kdde}

In this section, we establish the results needed for deriving the asymptotic mean integrated squared error (AMISE) and the AMISE-optimal smoothing parameter. Throughout this section, for a given derivative order $r$, we will employ the following assumptions to derive the asymptotic results.

\begin{enumerate}
\item[\refstepcounter{enumi}(A\number\value{enumi})\label{cond1}] The circular density $f$ is such that its derivative $f^{(r+2)}$ is continuous and square-integrable in $[-\pi,\pi)$.
\item[\refstepcounter{enumi}(A\number\value{enumi})\label{cond2}] The kernel $K$ is a bounded circular density, (reflectively) symmetric about zero, unimodal and its $r$-th derivative is square integrable in $[-\pi,\pi)$.
\item[\refstepcounter{enumi}(A\number\value{enumi})\label{cond3}] As $n\rightarrow\infty$, the value $h_{K}(\nu_n)=0$, $R_{K;r,2}(\nu_n)=\infty$, and $n^{-1} R_{K;r,2}(\nu_n)=0$. The function $R_{K;r,t}$, for $t\in\{1,2\}$, is defined as,
\begin{align*}
R_{K;r,t}(\nu)&=\left\{\begin{array}{ll}
(2 \pi)^{-1}\left(1+2 \sum_{j=1}^{\infty} \alpha_{K,j}^t(\nu) \right) & \text { if } r=0, \\
\mathfrak{s} \pi^{-1} \sum_{j=1}^{\infty} j^{t r} \alpha_{K,j}^t(\nu) & \text { otherwise, }
\end{array}\right. \\
\mathfrak{s} &=\left\{\begin{array}{ll}
-1 & \text { if } t=1 \text { and } r \text { modulo } 4 = 1, \\ 
-1 & \text { if } t=1 \text { and } r \text { modulo } 4 = 2, \\
1 & \text { otherwise. }
\end{array}\right.   
\end{align*}
\newcounter{enumTemp2}
\setcounter{enumTemp2}{\theenumi}
\end{enumerate}

Assumptions~\hyperref[cond1]{(A\ref{cond1})} and \hyperref[cond2]{(A\ref{cond2})} coincide with those employed in the standard linear case. For a large class of kernels, including the von Mises or the wrapped normal (see Section~\ref{kernel_choice}), we will see that Assumption~\hyperref[cond3]{(A\ref{cond3})} translates into the standard conditions on the bandwidth, namely, $h_{n}=0$ and $n h_{n}^{(2r+1)/2}=\infty$. The result in Theorem~\ref{AMISE_res} (see Section~\ref{proofs_ker_der} of the Appendix for a formal proof) states the AMISE order of the kernel derivative estimator \eqref{kernel_der}. If we also assume the following two extra conditions, we can derive an explicit expression of the AMISE.

\begin{enumerate}
\item[\refstepcounter{enumi}(E\number\value{enumi})\label{cond4}] $\lim_{n\rightarrow\infty} \int_{-\pi}^{\pi} \theta^4 K_{\nu_n}(\theta) d\theta=o(h_{n})$. 
\item[\refstepcounter{enumi}(E\number\value{enumi})\label{cond8}] $\lim_{n\rightarrow\infty} \int_{-\pi}^{\pi} \theta^2  (K_{\nu_n}^{(r)}\left(\theta \right))^2 d\theta = o[R_{K;r,2}(\nu_n)]$.
\newcounter{enumTemp}
\setcounter{enumTemp}{\theenumi}
\end{enumerate}

\begin{theorem}\label{AMISE_res}
Under the Assumptions~\hyperref[cond1]{(A\ref{cond1})}--\hyperref[cond3]{(A\ref{cond3})}, we have
\begin{equation*}
\mbox{AMISE}\left[ \hat{f}^{(r)}_\nu \right] = O\left(  h_{K}^2(\nu_n) \right)  +O\left( n^{-1} R_{K;r,2}(\nu_n) \right)
\end{equation*} 
If we also assume Conditions~\hyperref[cond4]{(E\ref{cond4})} and~\hyperref[cond8]{(E\ref{cond8})}, then the AMISE has the following explicit expression. 
\begin{equation}\label{amise_ker_der}
\mbox{AMISE}\left[ \hat{f}^{(r)}_\nu \right] = \frac{1}{4} h_{K}^2(\nu_n) \int_{-\pi}^{\pi} \left(f^{(r+2)} (\theta)\right)^2 d \theta + \frac{1}{n} R_{K;r,2}(\nu_n)
\end{equation}
\end{theorem}

Theorem~\ref{AMISE_res} states the asymptotic expression of the MISE which depend on the sample size $n$, the derivative of the density function $f^{(r+2)}$, the kernel $K$, and the smoothing parameter $\nu$. The complete expression of the AMSE is provided in Section~\ref{proofs_ker_der} of the Appendix.

\begin{remark}\label{remark_aprox_DM}
\cite{DiMarzio2009,DiMarzio2011} also analysed the AMISE of the circular kernel estimation obtaining a similar result as in~\eqref{amise_ker_der}, replacing $h_{K}(\nu_n)$ by $(1-\alpha_{K,2}(\nu_n))/2$. Note that, although not stated there, Condition~\hyperref[cond4]{(E\ref{cond4})} must be imposed to derive their asymptotic results. This can be seen in the remainder term of the asymptotic bias in the proof of Theorem~1 in \cite{DiMarzio2009}. Assuming Condition~\hyperref[cond4]{(E\ref{cond4})}, we obtain that $h_{K}(\nu_n)$ can be approximated by $(1-\alpha_{K,2}(\nu_n))/2$. Thus, asymptotically, both results coincide, but the expression provided in this paper will allow us to derive an expression of the optimal concentration. 
\end{remark}

The issue, under the general AMISE expression given in~\eqref{amise_ker_der}, is that it is not straightforward to know how to derive an explicit expression of the optimal smoothing parameter, unless a specific kernel, such as the von Mises is chosen \citep[see, e.g.,][]{DiMarzio2011}. This problem can be solved when the following extra condition is assumed.
\begin{enumerate}
\setcounter{enumi}{\theenumTemp}
\item[\refstepcounter{enumi}(E\number\value{enumi})\label{cond5}] As $n$ increases, $R_{K;r,2}(\nu_n)=Q_{K;r,2} h_{n}^{-(2r+1)/2}$, where $Q_{K;r,2}$ is a constant only depending on the kernel $K$ and $r$.
\setcounter{enumTemp}{\theenumi}
\end{enumerate} 
Note that under the Condition~\hyperref[cond5]{(E\ref{cond5})}, the Assumption~\hyperref[cond1]{(A\ref{cond3})} simplifies to $h_{n}=0$ and $n h_{n}^{(2r+1)/2}=\infty$. Using this last assumption, we can observe the classic \textit{variance-bias trade-off}, where the bias is reduced if $h_{n}$ is ``small'' and the variance decreases if $h_{n}$ is ``large''. The optimal smoothing parameter with respect to the AMISE criteria can be obtained using the following corollary of Theorem~\ref{AMISE_res}, which is a direct consequence of $h_{K}(\nu)>0$ (see Section~\ref{proofs_ker_der} of the Appendix). Note that, in the following corollary, we obtain that, under the previous assumptions, the AMISE order coincides with the one obtained in the linear case, $O(n^{-4/(2r+5)})$ \citep[][Section~2.5]{WandJones1995}.

\begin{corollary}\label{AMISE_cor}
Consider the Assumptions~\hyperref[cond1]{(A\ref{cond1})}--\hyperref[cond1]{(A\ref{cond3})} and~\hyperref[cond4]{(E\ref{cond4})}--\hyperref[cond5]{(E\ref{cond5})}. Then, for the kernel derivative estimator of order $r$ (see \eqref{kernel_der}), we have that, asymptotically, the optimal (AMISE) value of $h_{n}$ can be obtained from,
\begin{equation}\label{optimal_h}
h_{K; r; \mbox{\tiny{AMISE}}}= \left( \frac{(2r+1) Q_{K;r,2} }{n \int_{-\pi}^{\pi} \left( f^{(r+2)} (\theta) \right)^2 d \theta } \right)^{2/(2r+5)}.
\end{equation}
Under the previous assumptions, the minimal AMISE of $\hat{f}^{(r)}_\nu$ is equal to
\begin{equation}\label{AMISE_order}
\inf_{0\leq \nu <1} \mbox{AMISE}\left[ \hat{f}^{(r)}_\nu \right] =   \frac{2r+5}{8r+4}   \left( \frac{(2r+1) Q_{K;r,2} }{n \int_{-\pi}^{\pi} \left( f^{(r+2)} (\theta) \right)^2 d \theta } \right)^{4/(2r+5)} \int_{-\pi}^{\pi} \left(f^{(r+2)} (\theta)\right)^2 d \theta.
\end{equation}
\end{corollary}

The optimal concentration $\nu$ is the solution to the equation $h_{K}(\nu)=h_{K;r;\mbox{\tiny{AMISE}}}$, see Equation~\eqref{h_value}. Alternatively, to avoid the infinite sum in~\eqref{h_value}, one can also obtain the optimal $\nu$ by solving the equation $\alpha_{K,2}(\nu)=1-2h_{K;r;\mbox{\tiny{AMISE}}}$ (see Remark~\ref{remark_aprox_DM}).

As usual, when using data-based plug-in smoothing parameters, the main problem with employing the optimal $h_{K;r;\mbox{\tiny{AMISE}}}$ in Corollary~\ref{AMISE_cor}, is that its value depends on the unknown value of $\int_{-\pi}^{\pi} \left( f^{(r+2)} (\theta) \right)^2 d \theta$. A rule-of-thumb smoothing selector can be obtained by replacing $f$ with a simple and standard density function, such as the von Mises density. One can also follow the \cite{cwik1997} approach and replace $f$ with a mixture model, such as the mixture of von Mises. Both techniques were already proposed and studied in the circular literature when employing the von Mises density as the kernel $K$ \citep[see][]{Taylor2008,Oliveira2012}. Following \cite{sheather1991}, an alternative is to estimate density functionals related to $\int_{-\pi}^{\pi} \left( f^{(s)} (\theta) \right)^2 d \theta$. In the next section, we study how to obtain a kernel estimator of this last quantity, in the circular context. 

\section{Estimation of density functionals}\label{circular_functionals}

As mentioned in the previous section, an issue with using, in practice, the optimal smoothing parameter~\eqref{optimal_h} is its dependence on the unknown quantity $\int_{-\pi}^{\pi} \left( f^{(r+2)} (\theta) \right)^2 d \theta$. In this section, we will see how to estimate this last quantity using kernel techniques. For doing so, we first define the functional of the form

\begin{equation*}
\psi_s= \int_{-\pi}^{\pi}  f^{(s)} (\theta)  f (\theta) d\theta.
\end{equation*}

Note that under sufficient smoothness assumptions on $f$ (e.g., the needed conditions to apply integration by parts), we obtain that,

\begin{equation}\label{psi_derf}
\int_{-\pi}^{\pi} \left( f^{(s)} (\theta) \right)^2 d \theta = (-1)^s \psi_{2s}.
\end{equation}

Since $\psi_s =\mathbb{E} (f^{(s)} (\Theta))$, the following estimator can be employed to estimate the unknown quantity $\int_{-\pi}^{\pi} \left( f^{(r+2)} (\theta) \right)^2 d \theta$ on the optimal smoothing parameter~\eqref{optimal_h}, 

\begin{equation}\label{psi_estimator}
\hat \psi_{s;\rho} = \frac{1}{n} \overset {n} {\underset {i=1} \sum}  \hat{f}^{(s)}_\rho(\Theta_i)= \frac{1}{n^2} \overset {n} {\underset {i=1} \sum} \overset {n} {\underset {j=1} \sum} L_\rho^{(s)}\left(\Theta_i-\Theta_j\right),
\end{equation}
where $L$ and $\rho$ are a kernel and a concentration parameter, which are possibly different from $K$ and $\nu$.

Using the estimator~\eqref{psi_estimator}, a direct plug-in estimator of the smoothing parameter can be obtained from~\eqref{optimal_h}, replacing the quantity depending on the true $f$, by its estimator, in the following way,
 
\begin{equation}\label{plugin_h}
h_{K; r; \mbox{\tiny{PI}}}= \left( \frac{(2r+1) Q_{K;r,2} }{n (-1)^{r+2}\hat \psi_{2r+4; \rho} } \right)^{2/(2r+5)}.
\end{equation}

The problem with the direct plug-in estimator~\eqref{plugin_h}, is that it still depends on a choice of the pilot smoothing parameter $\rho$. Below, we establish the asymptotic theory to derive the optimal smoothing parameter of $\hat \psi_{2r+4; \rho}$. For obtaining that result, the following condition on the pilot parameter is required.

\begin{enumerate}
\setcounter{enumi}{3}
\item[\refstepcounter{enumi}(A\number\value{enumi})\label{cond6}] As $n\rightarrow\infty$, $R_{K;r,1}(\rho_n)=\infty$ and $n^{-1} R_{K;r,1}(\rho_n)=0$.
\end{enumerate}

Replacing, in Assumptions~\hyperref[cond1]{(A\ref{cond1})}--\hyperref[cond1]{(A\ref{cond3})}, the kernel $K$ by $L$, the smoothing parameter $\nu_n$ by $\rho_n$, and the order of the derivative $r$ by $s$; we obtain the following AMSE for the estimator $\hat \psi_{s; \rho}$ (see Section~\ref{proofs_functional} of the Appendix for a formal proof).

\begin{theorem}\label{amse_functional}
Under the Assumptions~\hyperref[cond1]{(A\ref{cond1})}--\hyperref[cond6]{(A\ref{cond6})},~\hyperref[cond4]{(E\ref{cond4})}, and~\hyperref[cond8]{(E\ref{cond8})} (using $L$, instead of $K$; $\rho_n$, instead $\nu_n$; and $s=r$, an even number); we have
\begin{align}\label{AMSE_psi}
\mbox{AMSE}\left[ \hat \psi_{s; \rho} \right] &= \left( n^{-1}  R_{L;s,1}(\rho_n) + \frac{1}{2} h_{L} (\rho_n) \psi_{s+2} \right)^2 
+ 2 n^{-2} \psi_{0} R_{L;s,2}(\rho_n)  \nonumber \\
& + 4 n^{-1} \left( \int_{-\pi}^{\pi} \left( f^{(s)} (\theta) \right)^2 f (\theta) d \theta - \psi_{s}^2 \right).
\end{align}
\end{theorem}

As for Corollary~\ref{AMISE_cor}, for deriving a simpler expression of the AMSE and the optimal value of $\mathfrak{h}_{n} \equiv h_{L} (\rho_n)$, we will assume the following extra conditions.

\begin{enumerate}
\setcounter{enumi}{\theenumTemp}
\item[\refstepcounter{enumi}(E\number\value{enumi})\label{cond7}] As $n$ increases, $R_{L;s,1}(\rho_n)=Q_{L;s,1} \mathfrak{h}_{n}^{-(s+1)/2}$, where $Q_{L;s,1}$ is a constant only depending on the kernel $L$ and the even number $s$. The sign of $Q_{L;s,1}$ is equal to the sign of $(-1)^{s/2}$.
\item[\refstepcounter{enumi}(E\number\value{enumi})\label{cond9}] $\lim_{n\rightarrow\infty} \int_{-\pi}^{\pi} \theta^4 L_{\rho_n}(\theta) d\theta=o(\mathfrak{h}_{n}^{5/4})$. 
\end{enumerate} 

From the AMSE expression in~\eqref{AMSE_psi}, we can see that the optimal $\mathfrak{h}_{n}$ value, in terms of AMSE, will depend on the relation between $R_{L;s,1}(\rho_n)$, $R_{L;s,2}(\rho_n)$, and $h_{L} (\rho_n)$. Conditions~\hyperref[cond5]{(E\ref{cond5})} and~\hyperref[cond7]{(E\ref{cond7})} help to establish this relation, from which the following corollary is derived (see Section~\ref{proofs_functional} of the Appendix).

\begin{corollary}\label{cor_optimal_h_psi}
Consider the assumptions of Theorem~\ref{amse_functional}, Conditions~\hyperref[cond5]{(E\ref{cond5})}, and~\hyperref[cond7]{(E\ref{cond7})}. Then, for the kernel estimator of $\psi_s$ (see \eqref{psi_estimator}), we have that, asymptotically, the optimal value, in terms of the AMSE expression, of ${\mathfrak{h}}_{n}$ can be obtained from,
\begin{equation}\label{optimal_h_psi}
\mathfrak{h}_{L; s; \mbox{\tiny{AMSE}}}= \left( -\frac{2  Q_{L;s,1} }{n  \psi_{s+2}} \right)^{2/(s+3)}.
\end{equation}
Under the previous assumptions, if Condition~\hyperref[cond9]{(E\ref{cond9})} is also assumed, the minimal AMSE for \eqref{psi_estimator} is of order $O(n^{-\min(5,s+3)/(s+3)})$. If $s$ is an even number greater than 2, the minimal AMSE would be equal to
\begin{equation}\label{amse_psi_1}
\inf_{0\leq \rho <1} \mbox{AMSE}\left[ \hat \psi_{s; \rho} \right] =  2  \psi_{0} Q_{K;r,2} \left( -\frac{\psi_{s+2} }{2  Q_{L;s,1}} \right)^{(2s+1)/(s+3)}  n^{-5/(s+3)}.  
\end{equation}
If $s=0$, the minimal AMSE is equal to
\begin{equation}\label{amse_psi_2}
\inf_{0\leq \rho <1} \mbox{AMSE}\left[ \hat \psi_{s; \rho} \right] =  4 \left( \int_{-\pi}^{\pi} \left( f^{(s)} (\theta) \right)^2 f (\theta) d \theta - \psi_{s}^2 \right) n^{-1} .
\end{equation}
When $s=2$, the minimal AMSE is equal to the sum of the right-hand sides of \eqref{amse_psi_1} and \eqref{amse_psi_2}.
\end{corollary}

Note that Condition~\hyperref[cond9]{(E\ref{cond9})} is only assumed to derive the same AMSE optimal order as in the linear case. If that condition is not fulfilled, then, when $s>0$, the leading term in the AMSE could be of a larger order (see Section~\ref{proofs_functional} of the Appendix). Another important consideration is that the sign of $\psi_{s+2}$ is the same as that of $(-1)^{s/2+1}$ and, using Condition~\hyperref[cond7]{(E\ref{cond7})}, it also coincides with the sign of $-Q_{L;s,1}$. Therefore, we always have that $\mathfrak{h}_{L; s; \mbox{\tiny{AMSE}}} \geq 0$ in \eqref{optimal_h_psi}. 

Using the quantity $\mathfrak{h}_{L; 2r+4; \mbox{\tiny{AMSE}}}$ in~\eqref{optimal_h_psi}, one can obtain the pilot smoothing parameter $\rho$ needed to derive the direct plug-in estimator~\eqref{plugin_h}. The issue, as in the linear case, is that $\mathfrak{h}_{L; 2r+4; \mbox{\tiny{AMSE}}}$ still depends on the unknown value of the functional $\psi_{2r+6}$. We comment on how to overcome this difficulty in Section~\ref{implementation_practice}.

\section{The kernel choice}\label{kernel_choice}

Theoretical results in Sections~\ref{asymptotic_kdde} and~\ref{circular_functionals} provide mathematical support for deriving the optimal smoothing parameter. However, we still did not discuss how to obtain the plug-in concentration parameter $\nu_{r; \mbox{\tiny{AMISE}}}$ from $h_{K; r; \mbox{\tiny{AMISE}}}$. As mentioned in Section~\ref{asymptotic_kdde}, we must solve the equation $h_{K}(\nu)=h_{K;r;\mbox{\tiny{AMISE}}}$. We also need to obtain the values of $Q_{K;r,1}$ and $Q_{K;r,2}$ in~\eqref{optimal_h} and \eqref{optimal_h_psi}. 

In this section, we study what happens when the ``most common'' circular models are employed as kernels. For doing so, we restrict our attention to the four standard choices of circular densities \citep[see][Section 3.5]{Mardia1999a}: cardioid, von Mises, wrapped normal, and wrapped Cauchy. Despite all of them satisfy Assumption~\hyperref[cond2]{(A\ref{cond2})}, the cardioid kernel, $K_{\tiny{\mbox{C}};\nu}(\theta)=(1+2\nu \cos(\theta))/(2\pi)$, with $|\nu|< 1/2$, does not meet Assumption~\hyperref[cond3]{(A\ref{cond3})} and will be hence discarded from the analysis. This can be seen by observing that, for any $\nu$, $h_{K_{\tiny{\mbox{C}}}}(\nu)= \pi^2/{3}-4 \nu \geq \pi^2/{3} -2$. The other three circular densities are studied in Sections~\ref{VMWNk} and \ref{WCk}.

Once the behaviour of the standard kernels is studied, a second question to discuss is which is the optimal kernel in the circular context. As in \cite{muller1984}, we study that optimality in terms of the AMISE expression in~\eqref{AMISE_order}. Fixing $f$ and the sample size $n$, in Section~\ref{wrapped_kernels}, we obtain the circular kernel that minimizes the AMISE.

\subsection{von Mises and wrapped normal kernels}\label{VMWNk}
First, denoting by $\mathcal{I}_j$ to the modified Bessel function of the first kind and order $j$, let us consider the density expressions of the von Mises (VM) and the wrapped normal (WN) kernels.

\begin{align*}
K_{\tiny{\mbox{VM}};\nu}(\theta) &= \frac{\exp\left( \kappa \, \cos(\theta) \right )}{\left ( 2 \pi \ {\mathcal{I}}_0(\kappa) \right)} = \frac{1}{2\pi} \left( 1+ 2\sum_{j=1}^{\infty} \frac{{\mathcal{I}}_j(\kappa)}{{\mathcal{I}}_0(\kappa)} \cos (j\theta) \right) \mbox{, where } \nu=\frac{{\mathcal{I}}_1(\kappa)}{{\mathcal{I}}_0(\kappa)} \mbox{, being } \kappa \geq 0.\\
K_{\tiny{\mbox{WN}};\nu}(\theta) &= \frac{1}{2\pi} \left( 1+ 2\sum_{j=1}^{\infty} \nu^{j^2} \cos ( j\theta) \right) \mbox{, with } \nu \in[0,1].
\end{align*}

For both kernels, if $\nu_n=1$ (equivalently $\kappa_n=\infty$, for the von Mises kernel), as $n\rightarrow\infty$, then $h_{K}(\nu_n)=0$. Even more, it is easy to show that, in that setting, Conditions~\hyperref[cond4]{(E\ref{cond4})} and~\hyperref[cond8]{(E\ref{cond8})} are satisfied. Thus, as $n\rightarrow\infty$, we obtain the following asymptotic approximations.

\begin{align}
h_{K_{\tiny{\mbox{VM}}}}(\nu_n) &= \frac{1}{2}\left(1- \frac{{\mathcal{I}}_2(\kappa_n)}{{\mathcal{I}}_0(\kappa_n)}\right) = \frac{1}{\kappa_n}\mbox{,} \quad \mbox{i.e.,} \quad  \kappa_n \overset{\tiny{\mbox{VM}}}{=} h_n^{-1}. \label{h_VM}\\
h_{K_{\tiny{\mbox{WN}}}}(\nu_n) &= \frac{1}{2}\left(1- \nu_n^4 \right)\mbox{,}  \quad \mbox{i.e., } \quad  \nu_n \overset{\tiny{\mbox{WN}}}{=} (1-h_n)^{1/4}.\label{h_WN}
\end{align}

From the previous equalities, we can see that $\kappa_n$ or $\nu_n$ are easily derived after computing the optimal/plug-in value of $h_n$ (see, e.g., \eqref{optimal_h} or \eqref{plugin_h}). Again, considering $\nu_n=1$, it can be seen that Conditions~\hyperref[cond5]{(E\ref{cond5})}--\hyperref[cond9]{(E\ref{cond9})} are satisfied for the von Mises and wrapped normal kernels. For both kernels and a non-negative integer number $r$, the values of the constants in \hyperref[cond5]{(E\ref{cond5})} and \hyperref[cond7]{(E\ref{cond7})} are the following ones,

\begin{equation}\label{Q_VM}
Q_{K;r,1}= (-1)^{r/2} \frac{r!}{2^{r/2} (r/2)! \sqrt{2 \pi}} \, (r \mbox{ being even}), \quad 
Q_{K;r,2}= \frac{(2r)!}{2^{2r+1} r! \pi^{1/2}}.
\end{equation}

We can see that the AMSE and AMISE results derived in Sections~\ref{asymptotic_kdde} and~\ref{circular_functionals} can be easily obtained in practice from these last quantities. These optimal asymptotic results will coincide for both, the von Mises and the wrapped normal kernel.

\subsection{Wrapped Cauchy kernel}\label{WCk}

The wrapped Cauchy kernel density expression is
\begin{equation*}
K_{\tiny{\mbox{WC}};\nu}(\theta) = \frac{1-\nu^2}{2\pi \left(1+\nu^2-2\nu\cos(\theta)  \right)} = \frac{1}{2\pi} \left( 1+ 2\sum_{j=1}^{\infty} \nu^j \cos (j\theta) \right) \mbox{, with } \nu \in[0,1].\\
\end{equation*}

From the last equality, we derive that $h_{K_{\tiny{\mbox{WC}}}}(\nu)=\pi^2/3+4 \text{Li}_2 (-\nu)$, where $\text{Li}_s$ is the polylogarithm of order $s$. The last implies $h_{K_{\tiny{\mbox{WC}}}}(\nu_n)=0$ if $\nu_n=1$ as $n\rightarrow\infty$. Also, the expressions of $R_{K_{\tiny{\mbox{WC}}};r,2}(\nu)$ can be obtained in terms of a polylogarithm,

\begin{align} \label{h_WC}
R_{K_{\tiny{\mbox{WC}}};r,2}(\nu)&=\left\{\begin{array}{ll}
(2 \pi)^{-1}\left(1+2 \text{Li}_{0} (\nu^2) \right) & \text { if } r=0, \\
\pi^{-1} \text{Li}_{-2r} (\nu^2)  & \text { otherwise, }
\end{array}\right\} =  O(h_n^{-(2r+1)}) .
\end{align}
 
If we consider a value of $h_n$ such that $h_{n}=0$ and $n h_{n}^{2r+1}=\infty$, Condition~\hyperref[cond3]{(A\ref{cond3})} is satisfied. Therefore, from Theorem~\ref{AMISE_res}, if also $f$ satisfies Assumption~\hyperref[cond1]{(A\ref{cond1})}, we obtain the following AMISE for the wrapped Cauchy kernel,
\begin{equation}\label{AMISE_WC}
\mbox{AMISE}\left[ \hat{f}^{(r)}_\nu \right] = O\left(  h_n^2 \right)  +O\left( n^{-1} h_n^{-(2r+1)} \right).
\end{equation}
The AMISE expression~\eqref{AMISE_WC} would be minimized if $h_n$ is of order $n^{-1/(2r+3)}$. Using the previous result, we obtain that the AMISE order of the wrapped Cauchy kernel is worse than that obtained with the von Mises or the wrapped normal kernel (see Corollary~\ref{AMISE_cor}). In particular, its minimal AMISE is equal to  
\begin{equation*}
\inf_{0\leq \nu <1} \mbox{AMISE}\left[ \hat{f}^{(r)}_\nu \right] =  O\left(  n^{-2/(2r+3)} \right) .
\end{equation*}

When searching for an explicit expression of the optimal smoothing parameter, one could be tempted to combine \eqref{amise_ker_der} and \eqref{h_WC}. But note that this cannot be done as Condition~\hyperref[cond4]{(E\ref{cond4})} is not verified. Thus, the explicit expression of the asymptotic bias cannot be obtained following the steps in Section~\ref{proofs_ker_der} of the Appendix. This means that Equations~\eqref{abias_ker_der} and, therefore, \eqref{amise_ker_der} are not necessarily true for this particular kernel. On the contrary, \cite{Tsuruta2017a} were able to obtain an optimal smoothing parameter for the wrapped Cauchy kernel from the results by \cite{DiMarzio2009}. Nevertheless, it should be noted that \cite{DiMarzio2009} results must not be employed for this kernel as Condition~\hyperref[cond4]{(E\ref{cond4})} is not satisfied (see Remark~\ref{remark_aprox_DM}).

\subsection{Wrapped bounded-support kernels}\label{wrapped_kernels}

To find the optimal kernels, consider a wrapped kernel $\displaystyle K_{\nu}(\theta) =\sum_{\ell=-\infty}^{\infty} K_{X; \lambda}(\theta + 2 \, \ell \, \pi )$, whose associated linear density $K_{X; \lambda}$ has bounded support, i.e., $K_{X; \lambda}(x)=0$ if $|x|> \lambda$. Then, $K_{\nu}(\theta) = K_{X; \lambda}(\theta )$, for all $\theta \in [-\pi,\pi)$, when $\lambda< \pi$. In that case,  $h= \int_{-\lambda}^{\lambda} x^2  K_{X; \lambda}(x) dx$ and $R_{K;r,2}(\nu)=\int_{-\lambda}^{\lambda}  (K_{X; \lambda}^{(r)}(x ) )^2 dx$. 

Consider the asymptotic case for the linear bounded-support kernels, i.e., $\lambda_n=0$. Fixing the unknown linear density function $f_X$, \cite{muller1984} gives explicit expressions of the bounded-support kernels minimizing the AMISE when estimating the density function and its derivatives, in the linear case. 

Assume now that the wrapped bounded-support kernel satisfies the Assumptions of Corollary~\ref{AMISE_cor}. Except for the values not depending on the kernel, the minimal AMISE of $\hat{f}^{(r)}_\nu$ in~\eqref{AMISE_order} is equal to the minimal AMISE obtained in the linear case. As a consequence, Theorem 2.4 of \cite{muller1984} can be employed to see that the optimal kernels in terms of AMISE coincide with the wrapped version of those employed on linear kernel estimation. 

The last means that the optimal kernel for circular density estimation is the wrapped Epanechnikov, whose associated linear density is $K_{X; \lambda}(x)=3(1-(x/\lambda)^2)/(4\lambda)$. When $\lambda < \pi$, the concentration is $\nu=(3\sin(\lambda)-3\lambda\cos(\lambda))/\lambda^3$, $h= \lambda^2/5$, and $R_{K;0,2}(\nu)= 3/(5\lambda)$. Thus, for the circular density estimation with the wrapped Epanechnikov, the optimal (AMISE) value of $h_n$ is obtained by taking $Q_{K;0,2}=3/(5\sqrt{5})$ in \eqref{optimal_h}. Given a circular density $f$, independently of the kernel, the minimal AMISE that could be obtained when estimating the circular density is equal to 

\begin{equation*}
\inf_{0\leq \nu <1} \mbox{AMISE}\left[ \hat{f}_\nu \right] =   \frac{5}{4}   \left( \frac{3}{5 \sqrt{5} n \int_{-\pi}^{\pi} \left( f^{(2)} (\theta) \right)^2 d \theta } \right)^{4/5} \int_{-\pi}^{\pi} \left(f^{(2)} (\theta)\right)^2 d \theta.
\end{equation*}

This is the AMISE obtained by the wrapped Epanechnikov. For the derivatives of the density, we refer to \cite{muller1984}. There, we can see, e.g., that the wrapped Biweight would be the optimal kernel for the first derivative of $f$. 

\section{The plug-in smoothing parameters}\label{implementation_practice}

In this section, we will study how to implement, in practice, the plug-in smoothing parameter $h_{K; r; \mbox{\tiny{PI}}}$ in~\eqref{plugin_h}. As mentioned in Section~\ref{circular_functionals}, the issue of directly employing~\eqref{plugin_h} is that the AMISE-optimal smoothing parameter for $\hat \psi_{s;\rho}$ will always depend on an unknown value of $\psi_{s+2}$. A way to overcome this difficulty is to provide an \textit{$l$-stage direct plug-in} smoothing selector. This procedure consists in estimate $\psi_{s}$ with $\hat \psi_{s+2;\rho}$, in an iterative process, until some point in which we replace $\psi_{2r+2l+4}$ by its value obtained with a simple density (see Section~\ref{algorithm_two_stage} for more details). An alternative selector can be computed by noting that the smoothing parameter for $\hat{f}^{(r)}$ can be obtained as a function of the smoothing parameter for $\hat \psi_{2r+4}$. This allows us to construct a \textit{solve-the-equation} rule. In Section~\ref{algorithm_solve_equation}, we discuss in more detail how this selector is derived.

\subsection{The two-stage direct plug-in smoothing selector}\label{algorithm_two_stage}

The procedure to obtain the $l$-stage direct plug-in smoothing selector consists in, at stage 0, using a simple rule of thumb to compute the smoothing parameter of $\hat \psi_{2r+4+2l;\rho}$. Once this initial step is achieved, from that estimator of $\psi_{2r+4+2l}$, the following functional estimators are derived in an iterative process (see Algorithm~\ref{algorithm_proposed} for details).

Two decisions remain from this brief explanation: the number of stages $l$ and which reference density should be employed at stage 0. Regarding the number of stages $l$, we suggest employing $l=2$ for two reasons. First, since $l=2$ is a common choice in the linear case \citep[see, e.g.,][Section 3.6]{WandJones1995}. A second reason was obtained when replicating the simulation study in Section~\ref{simulation_study}, with $l=3$. In that case, similar results were obtained with respect to those obtained with $l=2$, with the added inconvenience of the extra computational time.

As mentioned before, at stage 0, a reference density is needed to compute the smoothing parameter of $\hat \psi_{2r+2l+4;\rho}$. Here, in the same spirit of the original rule of thumb proposed by \cite{Taylor2008}, a natural selection would be to replace $\psi_{2r+2l+4;\rho}$ in~\eqref{optimal_h_psi} by the quantity obtained when assuming that the true density follows a von Mises distribution. In the circular case, the issue of employing that simple strategy is that a uniform estimation of the density can be obtained even if the true distribution is not uniform. This occurs, for example, when considering distributions with antipodal symmetry \citep[see also][for further discussion on this topic]{Oliveira2012}. The consequence would be to have a value of $\hat \psi_{2r+2l+4;\rho}$ close to zero, which derives in a ``large'' value of the smoothing parameter at the next stage $\mathfrak{h}_{L; 2r+2l+2; \mbox{\tiny{PI}}}$.

To avoid that last issue, while still having a simple model, one can use as the reference density the following mixture of $M$ von Mises, all having the same concentration parameter $\kappa \geq 0$.
\begin{equation}\label{mixture_von_Mises}
f_{\tiny{\mbox{MvM}}}(\theta;M;\bm{\mu},\kappa,\bm{w}) = \frac{1}{\left ( 2 \pi \ {\mathcal{I}}_0(\kappa) \right)} \sum_{m=1}^{M} w_m \exp\left( \kappa \, \cos(\theta-\mu_m) \right ),
\end{equation} 
where the parameters $\mu_m \in [-\pi,\pi)$, $w_m \in [0,1]$, for all $m \in\{1,\ldots, M\}$, and with $\sum_{m=1}^{M} w_m=1$. The value of $\psi_{2r+2l+4;\rho}$ is calculated from the density~\eqref{mixture_von_Mises}, replacing its parameters by their maximum likelihood estimates obtained from the sample. An algorithm providing the maximum likelihood estimates for the density~\eqref{mixture_von_Mises} was implemented in the R \citep{R21} library \texttt{movMF} by \cite{Hornik2014}. In practice, following \cite{Oliveira2012}, the value of $M$ can be chosen, using the Akaike Information Criterion (AIC), by comparing the results obtained with the mixtures~\eqref{mixture_von_Mises} of $M=1,\ldots,M_{\max}$ components.

In Algorithm~\ref{algorithm_proposed}, we summarize the steps that are needed to obtain our proposed two-stage direct plug-in smoothing parameter. There, for simplicity, we use $K=L$, a kernel that satisfies Assumptions~\hyperref[cond1]{(A\ref{cond1})}--\hyperref[cond1]{(A\ref{cond6})} and the extra Conditions~\hyperref[cond4]{(E\ref{cond4})}--\hyperref[cond7]{(E\ref{cond7})}.

\begin{algorithm}\label{algorithm_proposed}
Two-stage direct plug-in smoothing selector.

\begin{steps}
\item Use a rule of thumb to obtain the estimator of $\psi_{2r+8}$, $\hat \psi_{2r+8;\tiny{\mbox{RT}}}$. This can be achieved as follows. 
  \begin{substeps}
  \item For every $M=1,\ldots,M_{\max}$; obtain the maximum likelihood estimators of the parameters in the mixture of $M$ von Mises, all having the same concentration parameter, i.e., of $\bm{\mu}$, $\kappa$, and $\bm{w}$ in~\eqref{mixture_von_Mises}.
	\item Select the number of components in the mixture, $M_{\tiny{\mbox{AIC}}}$, employing the AIC.
  \item Using the density $f_{\tiny{\mbox{MvM}}}(\cdot;M_{\tiny{\mbox{AIC}}};\hat{\bm{\mu}},\hat \kappa,\hat{\bm{p}})$, compute the value $\hat \psi_{2r+8;\tiny{RT}}$ (see Equality~\eqref{psi_derf}).
  \end{substeps}
\item Estimate $\psi_{2r+6}$ using the estimator $\hat \psi_{2r+6;\rho_1}$, where $\rho_1$ is the plug-in concentration parameter relying on $\hat \psi_{2r+8;\tiny{\mbox{RT}}}$.
\begin{substeps}
\item Obtain $\mathfrak{h}_{K; 2r+6; \mbox{\tiny{PI}}}$ from~\eqref{optimal_h_psi}, replacing $\psi_{2r+8}$ by $\hat \psi_{2r+8;\tiny{\mbox{RT}}}$. 
\item The value $\rho_1$ is the one that satisfies $h_{K}(\rho_1)=\mathfrak{h}_{K; 2r+6; \mbox{\tiny{PI}}}$, see Equation~\eqref{h_value}.
\end{substeps}
\item Estimate $\psi_{2r+4}$ using the estimator $\hat \psi_{2r+4;\rho_2}$, where $\rho_2$ is the plug-in concentration parameter relying on $\hat \psi_{2r+6;\rho_1}$.
\item Compute the AMISE-optimal smoothing parameter for $\hat{f}^{(r)}$, relying on $\hat \psi_{2r+4;\rho_2}$.
\begin{substeps}
\item Obtain the two-stage direct plug-in smoothing parameter $h_{K; r; \mbox{\tiny{PI}}}$ from~\eqref{plugin_h}, with the pilot smoothing parameter $\rho_2$.
\item The selected concentration parameter $\nu_{ \mbox{\tiny{DPI}}}$ is obtained from the value that satisfies $h_{K}(\nu_{ \mbox{\tiny{DPI}}})=h_{K; r; \mbox{\tiny{PI}}}$.
\end{substeps}
\end{steps}
\end{algorithm}

The uniform distribution belongs to all the reference distributions mentioned before. Thus, in practice, the denominator of~\eqref{plugin_h} or~\eqref{optimal_h_psi} could be equal to zero. If that occurs, we suggest directly returning a value of the concentration parameter $\nu_{ \mbox{\tiny{DPI}}}=0$, which would correspond with the uniform estimation of the density. 

Given a finite value of $n$ in~\eqref{plugin_h} or~\eqref{optimal_h_psi}, the equation $h_{K}(\nu)=h_{K;r;\mbox{\tiny{PI}}}$, or its approximation $\alpha_{K,2}(\nu)=1-2h_{K;r;\mbox{\tiny{AMISE}}}$, cannot be solved for ``large'' values of $h_{K;r;\mbox{\tiny{PI}}}$. The reason is that $h_{K}(\nu)$ and $\alpha_{K,2}(\nu)$ are non-negative and bounded for any value of $\nu$. Since a ``large'' value of $h_{K}(\nu)$ corresponds to $\nu=0$, if the equation $h_{K}(\nu)=h_{K;r;\mbox{\tiny{AMISE}}}$ cannot be solved, we also suggest to employ the uniform as the density estimator.

\subsection{Solve-the-equation plug-in smoothing selector}\label{algorithm_solve_equation}

An alternative to the previous smoothing selector is the solve-the-equation rule. This rule consists in searching for the smoothing parameter that satisfies

\begin{equation}\label{solve_h}
h_{K; r; \mbox{\tiny{STE}}}= \left( \frac{(2r+1) Q_{K;r,2} }{n (-1)^{r+2}\hat \psi_{2r+4; \rho_{K; r; \mbox{\tiny{STE}}}} } \right)^{2/(2r+5)} \quad \mbox{ with } h_{K}(\rho_{K; r; \mbox{\tiny{STE}}})=\gamma(h_{K; r; \mbox{\tiny{STE}}}).
\end{equation}

Now, the smoothing parameter for $\hat \psi_{2r+4}$ is a function of the smoothing parameter for $\hat{f}^{(r)}$. We suggest taking the function $\gamma$ by looking at the relation between the optimal smoothing parameter of these two estimators.

\begin{equation*}
\mathfrak{h}_{L; 2r+4; \mbox{\tiny{AMSE}}}= \left( (-1)^{r+1}\frac{2  Q_{L;2r+4,1} }{(2r+1) Q_{K;r,2}} \right)^{2/(2r+7)} \left( \psi_{2r+4}/\psi_{2r+6} \right)^{2/(2r+7)} h_{K; r; \mbox{\tiny{AMISE}}}^{(2r+5)/(2r+7)}.
\end{equation*}

Using a plug-in rule, this last relation suggests taking the following function,
\begin{equation}\label{gammaf}
\gamma\left({h} \right)= \left( (-1)^{r+1}\frac{2  Q_{L;2r+4,1} }{(2r+1) Q_{K;r,2}} \right)^{2/(2r+7)} \left( \hat \psi_{2r+4;\rho_1}/\hat \psi_{2r+6;\rho_2} \right)^{2/(2r+7)} h^{(2r+5)/(2r+7)}.
\end{equation}

The two concentration parameters of the density functional estimators, $\hat \psi_{2r+4}$ and $\hat \psi_{2r+6}$, can be obtained from~\eqref{optimal_h_psi}. Those smoothing parameters will depend on some other density functionals, $\psi_{2r+6}$ and $\psi_{2r+8}$. Following \cite{sheather1991}, we suggest estimating these two new functionals with a rule of thumb. In Algorithm~\ref{algorithm_proposed2}, we summarize the steps that are needed to obtain our proposed solve-the-equation plug-in smoothing parameter. Again, for simplicity, we use $K=L$, a kernel that satisfies Assumptions~\hyperref[cond1]{(A\ref{cond1})}--\hyperref[cond1]{(A\ref{cond6})} and the extra Conditions~\hyperref[cond4]{(E\ref{cond4})}--\hyperref[cond7]{(E\ref{cond7})}.

\begin{algorithm}\label{algorithm_proposed2}
Solve-the-equation plug-in smoothing selector.

\begin{steps}
\item Use a rule of thumb to obtain the estimator of $\psi_{2r+6}$ and $\psi_{2r+8}$, namely $\hat \psi_{2r+6;\tiny{\mbox{RT}}}$ and $\hat \psi_{2r+8;\tiny{\mbox{RT}}}$. This can be done as in Step 1 of Algorithm~\ref{algorithm_proposed}. 
\item Estimate $\psi_{2r+4}$ and $\psi_{2r+6}$ using the estimators $\hat \psi_{2r+4;\rho_1}$ and $\hat \psi_{2r+6;\rho_2}$, where $\rho_1$ is the plug-in concentration parameter relying on $\hat \psi_{2r+6;\tiny{\mbox{RT}}}$, and $\rho_2$ the one relying on $\hat \psi_{2r+8;\tiny{\mbox{RT}}}$.
\item Employ the estimators $\hat \psi_{2r+4;\rho_1}$ and $\hat \psi_{2r+6;\rho_2}$ to obtain the function $\gamma$ in \eqref{gammaf}.
\item Using the $\gamma$ function of Step 3, select the value of $h_{K; r; \mbox{\tiny{STE}}}$ by solving Equation~\eqref{solve_h}. The solve-the-equation concentration parameter $\nu_{ \mbox{\tiny{STE}}}$ is obtained from the value that satisfies $h_{K}(\nu_{ \mbox{\tiny{STE}}})=h_{K; r; \mbox{\tiny{STE}}}$.
\end{steps}
\end{algorithm}

Note that two extra functional estimators must be estimated in Step 1 of Algorithm~\ref{algorithm_proposed2}. For this reason, the obtained smoothing parameter could be considered as a \textit{two-stages solve-the-equation plug-in selector}. As for the direct rule, the number of stages could be increased to estimate the density functionals. In particular, a rule of thumb can be employed to estimate $\psi_{2r+10}$ or $\psi_{2r+12}$ and then, in an iterative process, the values of $\rho_1$ and $\rho_2$ are obtained (as in Steps 2 and 3 of Algorithm~\ref{algorithm_proposed}). The simulation study in Section~\ref{simulation_study} was carried out under the same conditions, using three (relying on $\hat \psi_{2r+8;\tiny{\mbox{RT}}}$ and $\hat \psi_{2r+10;\tiny{\mbox{RT}}}$) and four (relying on $\hat \psi_{2r+10;\tiny{\mbox{RT}}}$ and $\hat \psi_{2r+12;\tiny{\mbox{RT}}}$) stages. The results in practice were similar to those obtained with Algorithm~\ref{algorithm_proposed2}, with the drawback of the extra computational time.

\section{Simulation study}\label{simulation_study}

In this section, we performed a simulation study to analyse the performance of the direct two-stage plug-in concentration parameter (see Algorithm~\ref{algorithm_proposed}) and the solve-the-equation plug-in smoothing selector (see Algorithm~\ref{algorithm_proposed2}). We focused only on the density estimation case ($r=0$ in \eqref{kernel_der}), with $K$ being the von Mises kernel. The reason is that their effectiveness can be compared with other smoothing parameters proposed in the literature. But note that our data-driven smoothing parameters could be employed to estimate any derivative and with other kernels. In particular, in this framework, slightly better results are expected by employing the wrapped Epanechnikov kernel (see Section~\ref{wrapped_kernels}).

Regarding the choice of $M_{\max}$ at stage 0 (see Step 1 of Algorithms~\ref{algorithm_proposed} and~\ref{algorithm_proposed2}), we studied two scenarios. First, the reference model for the plug-in smoothing parameters was a simple von Mises ($M_{\max}=1$). As a second choice, we allowed for mixture models~\eqref{mixture_von_Mises}, with $M_{\max}=5$, in the reference density. In Tables~\ref{table_results1} and~\ref{table_results2}, we show both results, $M_{\max}=1$ $(\nu_{ \mbox{\tiny{DPI}};1})$ and $M_{\max}=5$ $(\nu_{ \mbox{\tiny{DPI}};5})$, for the direct plug-in concentration parameter. While for the solve-the-equation plug-in smoothing selector $(\nu_{ \mbox{\tiny{STE}};1})$, we only show the results when $M_{\max}=1$, given that our empirical experiments reveal that for the shown ``small/moderate'' sample sizes, most of the time $M_{\max}=1$ performs better, even in the more complex models, and it requires less computational time.

The results of using a common concentration parameter in the reference density (see~\eqref{mixture_von_Mises}) at Step 1 were compared with those of the full von Mises mixture (allowing for different concentration parameters in each component). For most of the studied scenarios, better or similar results were obtained using a common value of $\kappa$, while keeping the computational efficiency. For that reason, we only show the results obtained using a common concentration parameter (as described in Algorithms~\ref{algorithm_proposed} and~\ref{algorithm_proposed2}). 

We will compare the performance of the proposed concentration parameters with the following three rules for smoothing selection in circular density estimation, all implemented in the R library \texttt{NPCirc} \citep{Oliveira2014b}. 

\begin{itemize}
\item Rule of thumb of \cite{Taylor2008}, $\nu_{ \mbox{\tiny{RT}}}$, where to compute $\psi_{4}$ in \eqref{plugin_h}, it is assumed that $f$ follows a von Mises.
\item Plug-in rule of \cite{Oliveira2012}, $\nu_{ \mbox{\tiny{MvM}}}$, where to compute $\psi_{4}$ in \eqref{plugin_h}, it is assumed that $f$ follows a mixture of $M$ von Mises, with different concentration parameters. The number $M$ is chosen between 1 and 5, according to the AIC.
\item Likelihood cross-validation of \cite{Hall1987}, $\nu_{ \mbox{\tiny{LCV}}}$. Consider the leave-one-out estimator $\hat{f}_{-i;\nu}(\theta)$, which corresponds with the kernel density estimation~\eqref{kernel}, leaving out the $i$-th observation. Then, the concentration parameter $\nu_{ \mbox{\tiny{LCV}}}$ is obtained from the value of $\nu$ that maximizes $\prod_{i=1}^n \hat{f}_{-i;\nu}(\Theta_i)$.
\end{itemize}

The reasons for not showing the results with the other two well-known rules for obtaining the concentration parameter can be found in \cite{Oliveira2012}. \cite{Oliveira2012} mentioned that their empirical experiments show that the likelihood cross-validation rule provides a more stable behaviour than the least-squares cross-validation method \citep{Hall1987}. The reason for not including the bootstrap smoothing selector of \cite{DiMarzio2011} is that it relies on an optimization algorithm that searches for a local minimum. In that optimization, it is needed to impose the possible range of concentration values and, when $n$ is not ``too large'' ($n=100$), the needed local minimum was not found for several samples. For $n$ ``large'' ($n=250$ or $n=500$), \cite{Oliveira2012} show that their plug-in rule outperforms the bootstrap method except when data is generated from the simplest models (M1--M4 below).

For comparing the performance of the different rules, we have employed the 20 reference models that can be found in \cite{Oliveira2012}. They correspond to the uniform (M1), simple unimodal models (symmetric, M2--M5, and asymmetric M6), two-component mixture models (M7--M10), models with more than two components (M11--M16), and more complex models (M17--M20). Their density representations can be found in Tables~\ref{table_results1} and~\ref{table_results2}, and their density expressions appear in \cite{Oliveira2012}. From each model, we have generated 1000 random samples of sizes $n=50$ and $n=100$. Given the true density $f$, for each sample, we have computed the integrated squared error (ISE) as the criterion to analyse the performance of the different estimators.

\begin{equation}\label{ISE_nu}
\mbox{ISE}(\nu)= \int_{-\pi}^{\pi}{\left( \hat{f}_\nu(\theta) - f(\theta) \right)^2 d\theta} .
\end{equation}

\begin{table}
\centering
\scalebox{0.77}{
\begin{tabular}{llllllll}
  \hline
Model & M1 & M2 & M3 & M4 & M5 & M6 & M7 \\ 
$n=50$ & \includegraphics[width=20mm]{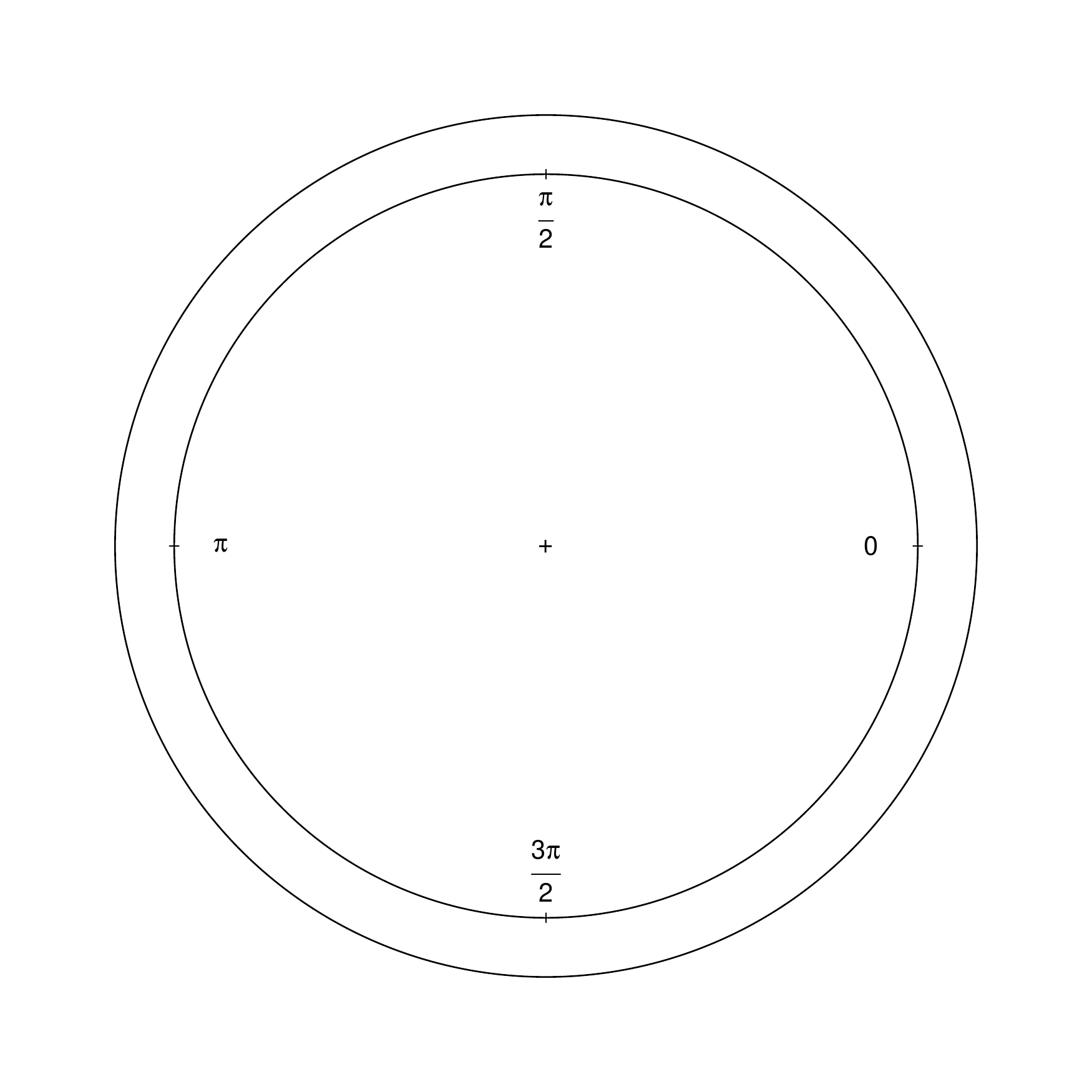} & \includegraphics[width=20mm]{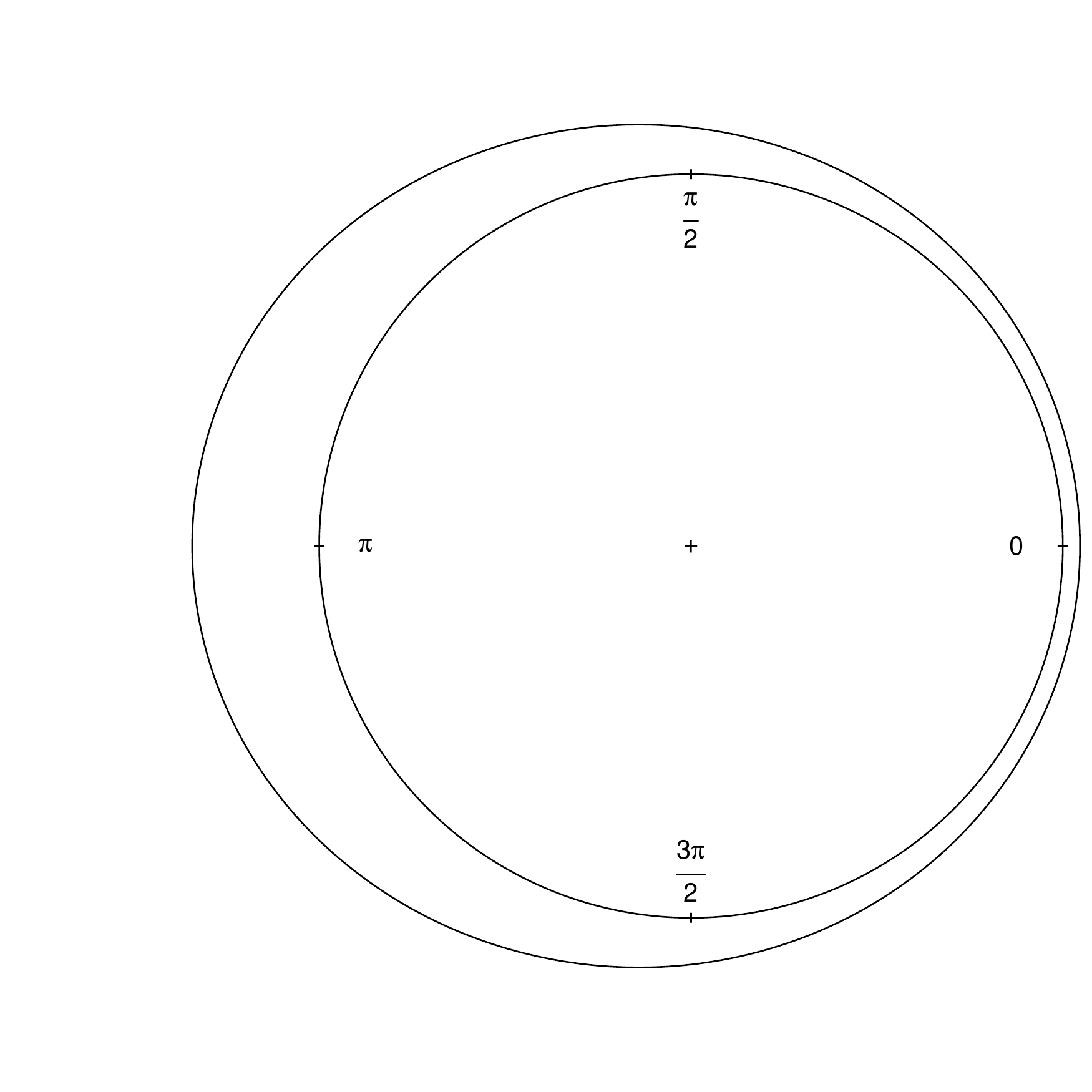}& \includegraphics[width=20mm]{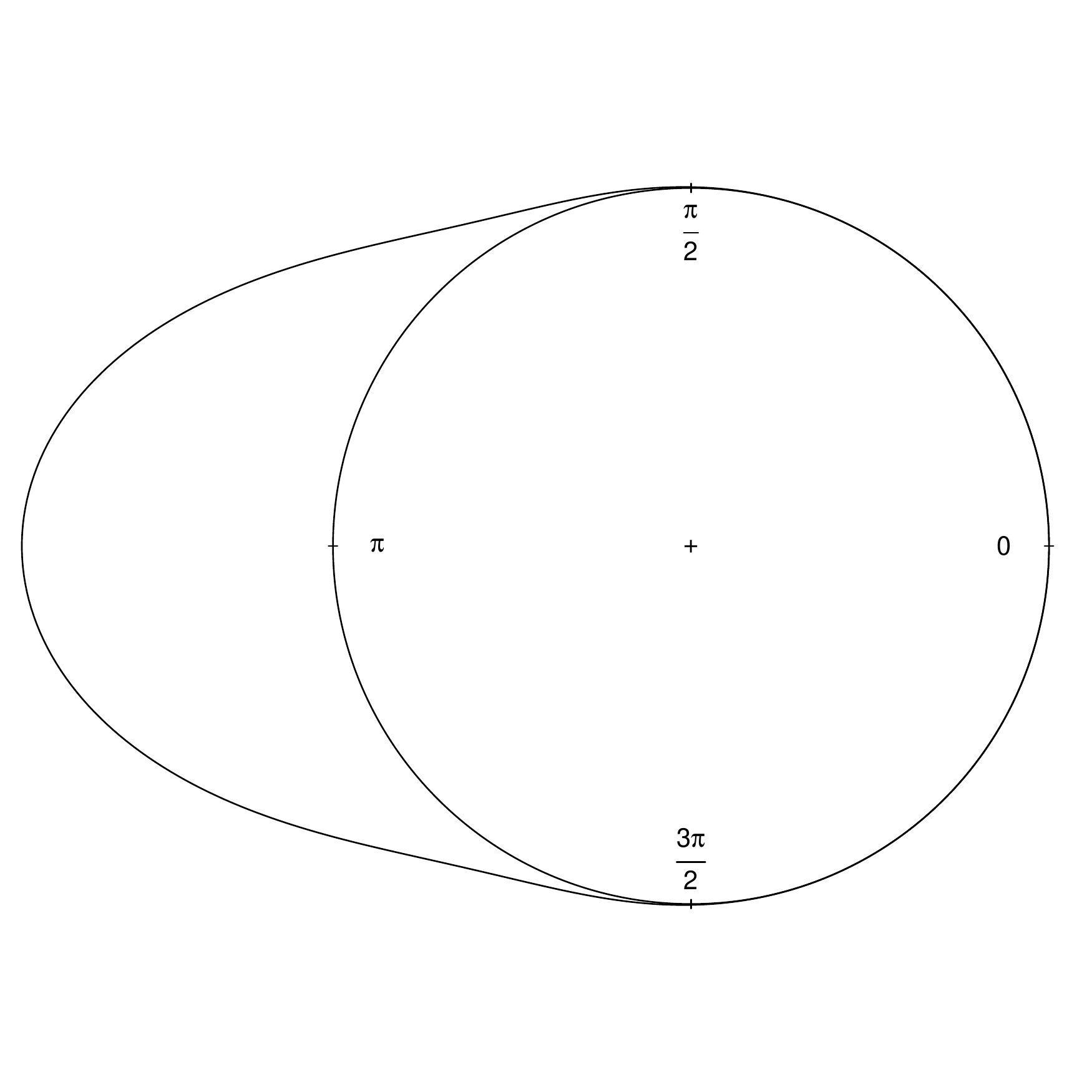}& \includegraphics[width=20mm]{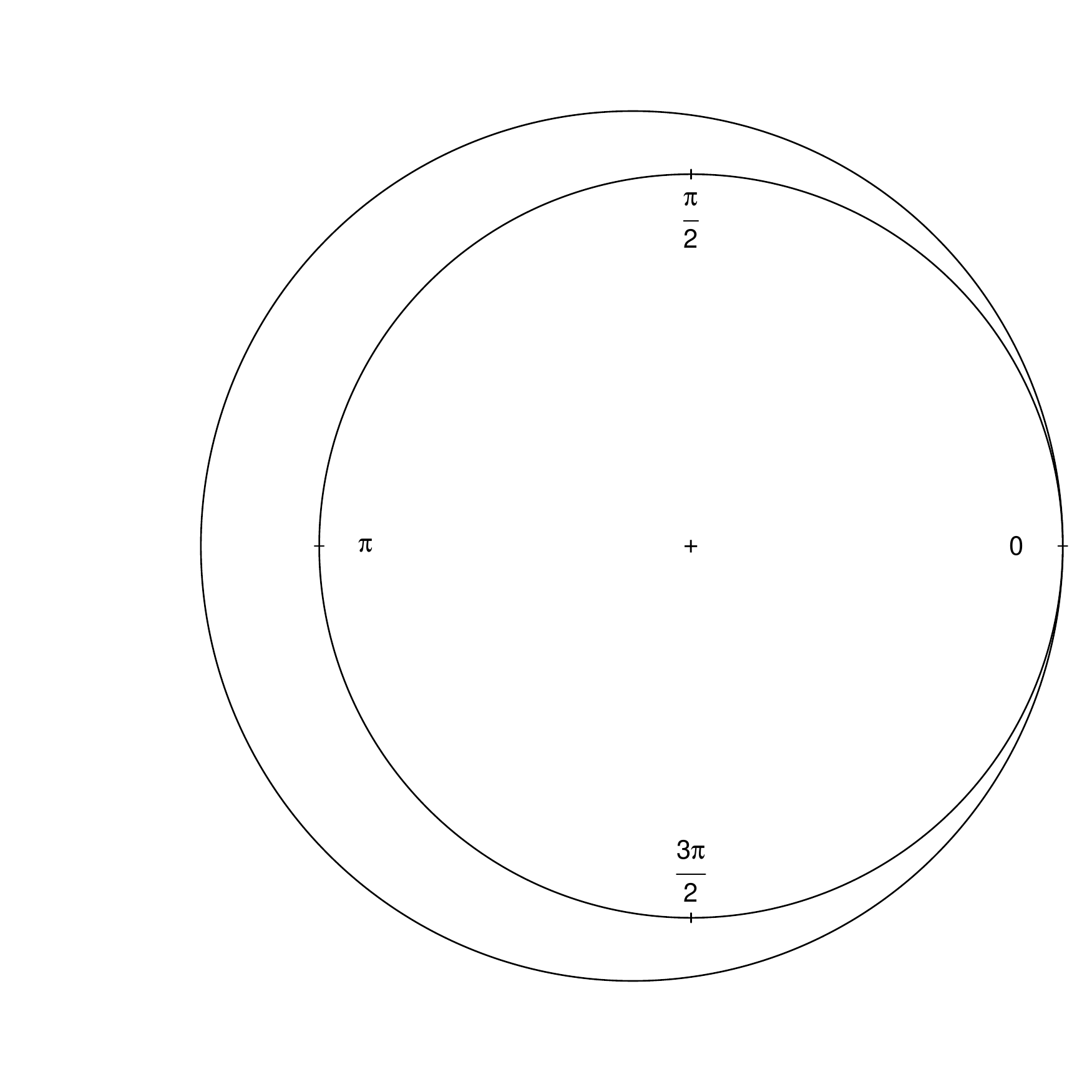}& \includegraphics[width=20mm]{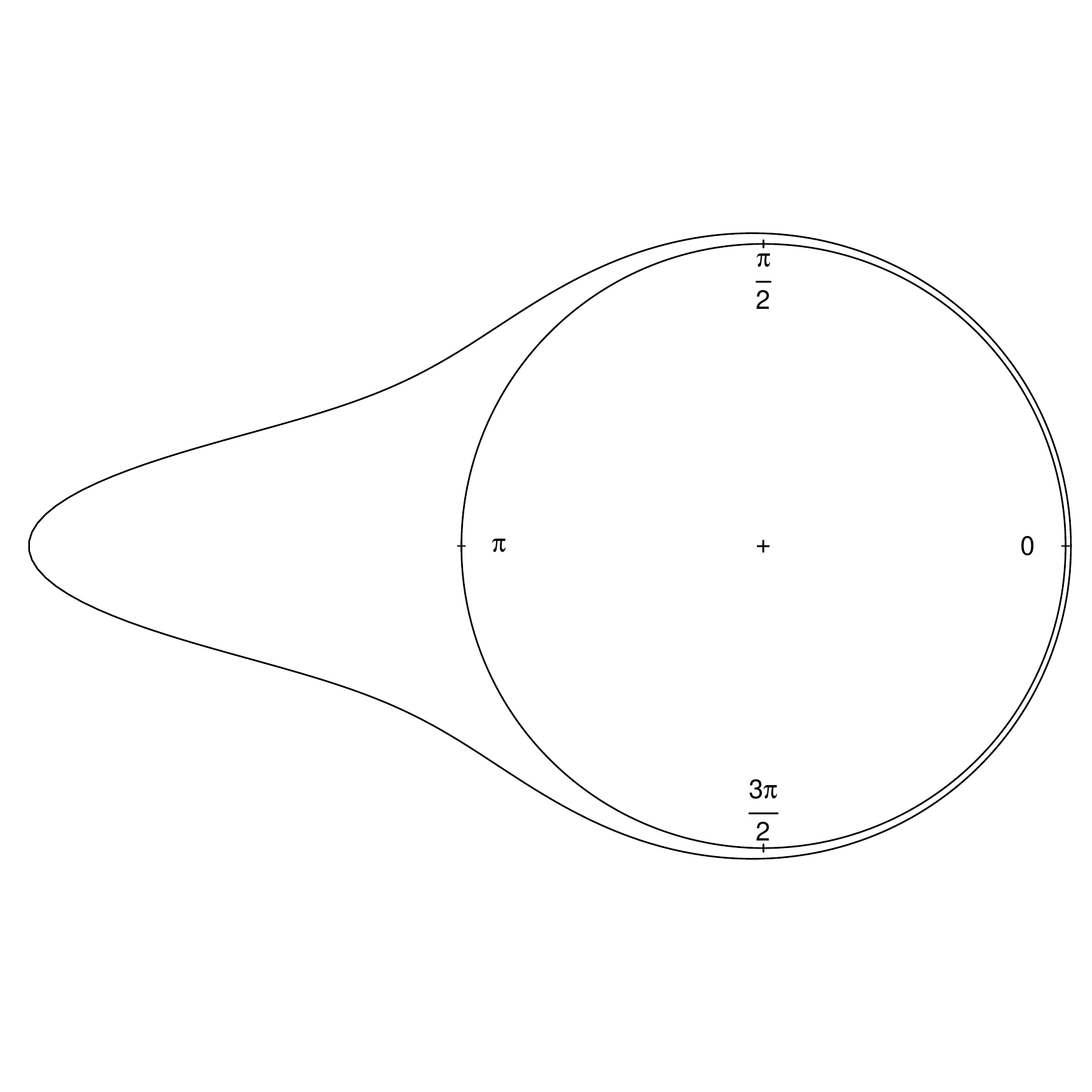}& \includegraphics[width=20mm]{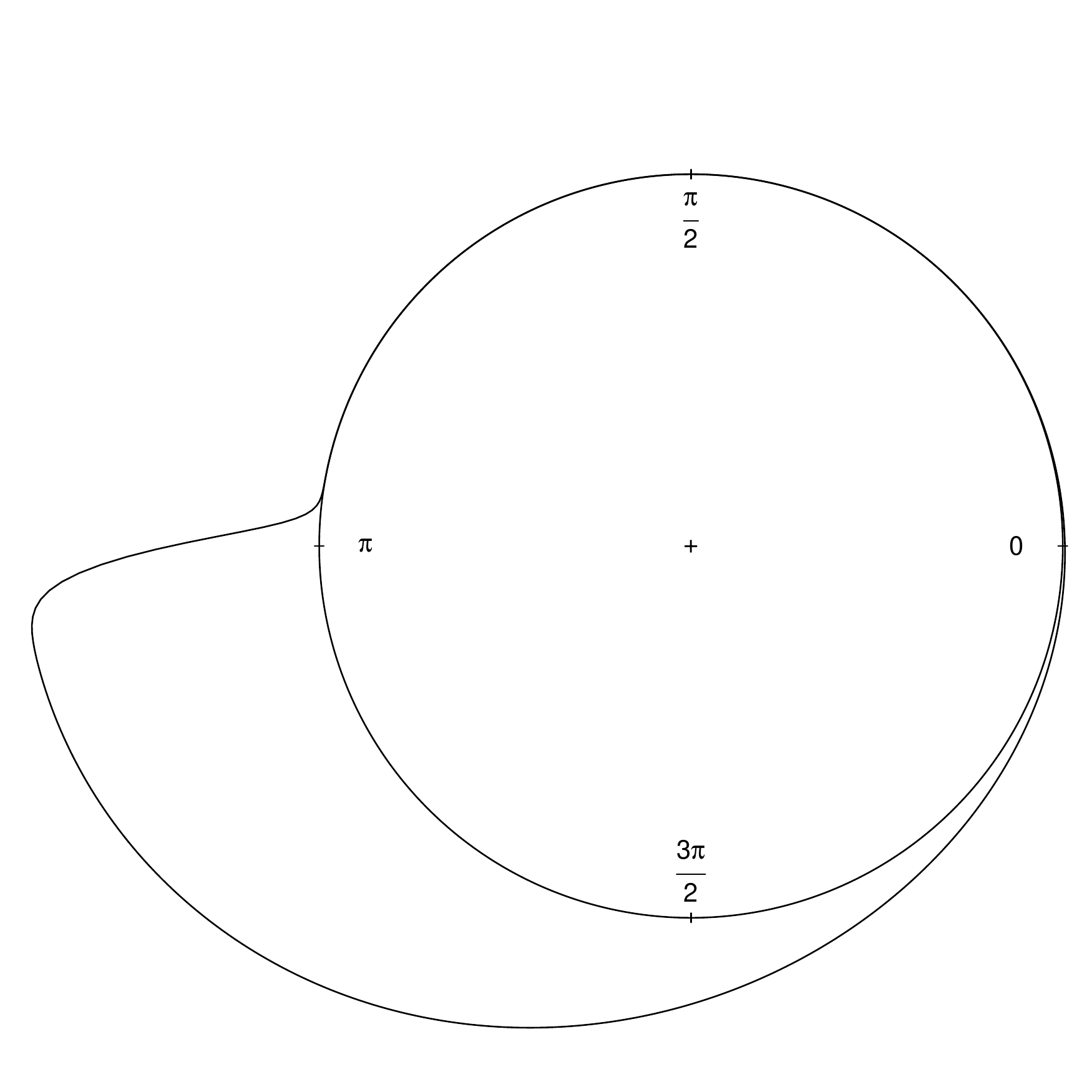}& \includegraphics[width=20mm]{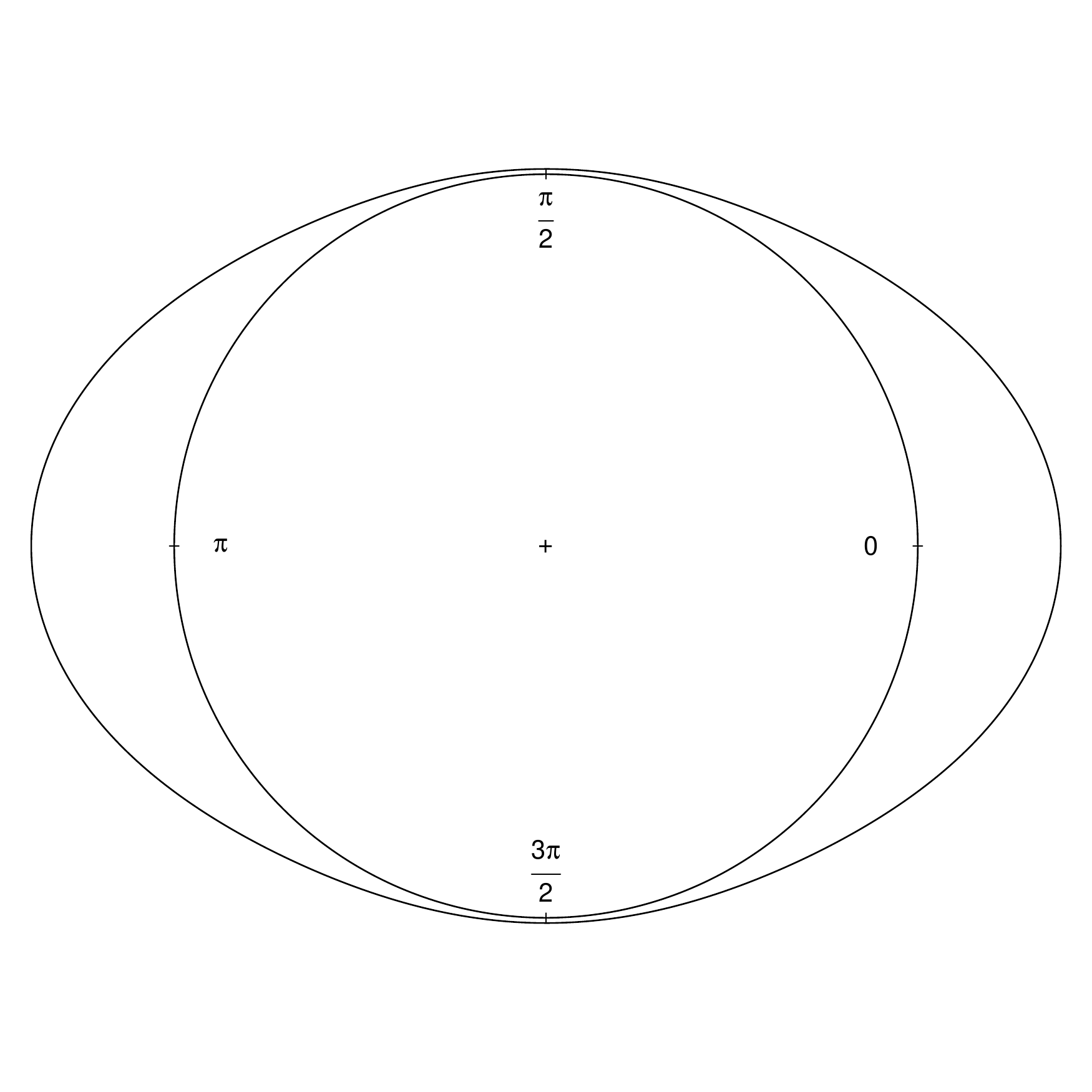} \\
$\nu_{ \mbox{\tiny{GS}}}$ & 0 (0) & 0.762 (0.644) & 1.67 (1.346) & 0.695 (0.54) & 4.199 (2.499) & 3.258 (1.433) & 1.758 (1.062) \\ 
$\nu_{ \mbox{\tiny{DPI}};1}$  & \textbf{\textcolor[rgb]{0.3,0.3,0.3}{0.381 (0.51)}} & \textbf{1.068 (0.813)} & \textbf{\textcolor[rgb]{0.3,0.3,0.3}{2.283 (1.685)}} & \textbf{\textcolor[rgb]{0.3,0.3,0.3}{1.04 (0.753)}} & 5.495 (3.124) & \textbf{3.766 (1.61)} & 2.503 (1.091) \\ 
$\nu_{ \mbox{\tiny{DPI}};5}$ &  0.473 (0.746) & \textbf{\textcolor[rgb]{0.3,0.3,0.3}{1.134 (0.938)}} & 2.498 (2.289) & 1.317 (1.191) & \textbf{\textcolor[rgb]{0.3,0.3,0.3}{5.345 (3.262)}} & 4.645 (2.616) & 2.177 (1.399) \\ 
$\nu_{ \mbox{\tiny{STE}};1}$ & 0.847 (0.83) & 1.319 (0.999) & 2.586 (2.074) & 1.319 (1.005) & \textbf{5.105 (3.016)} & \textbf{\textcolor[rgb]{0.3,0.3,0.3}{3.883 (1.832)}} & \textbf{1.965 (1.166)} \\ 
$\nu_{ \mbox{\tiny{RT}}}$ & \textbf{0.064 (0.176)} & 1.141 (0.958) & \textbf{1.98 (1.435)} & \textbf{0.93 (0.7)} & 10.893 (4.138) & 4.156 (1.469) & 10.471 (0.481) \\ 
$\nu_{ \mbox{\tiny{MvM}}}$ & 2.052 (2.807) & 2.688 (2.65) & 3.795 (3.813) & 2.646 (2.743) & 5.51 (3.244) & 5.57 (3.402) & 2.931 (2.67) \\ 
$\nu_{ \mbox{\tiny{LCV}}}$ & 0.595 (1.139) & 1.228 (1.041) & 2.45 (2.039) & 1.175 (1.051) & 11.45 (4.434) & 4.201 (1.952) & \textbf{\textcolor[rgb]{0.3,0.3,0.3}{2.086 (1.261)}} \\ 
   \hdashline
$n=100$ & &&&&&&\\ 
	$\nu_{ \mbox{\tiny{GS}}}$ & 0 (0) & 0.467 (0.363) & 1.1 (0.868) & 0.442 (0.308) & 2.648 (1.507) & 2.272 (0.968) & 1.076 (0.572) \\ 
$\nu_{ \mbox{\tiny{DPI}};1}$  & \textbf{\textcolor[rgb]{0.3,0.3,0.3}{0.165 (0.226)}} & \textbf{0.609 (0.414)} & \textbf{\textcolor[rgb]{0.3,0.3,0.3}{1.412 (1.012)}} & \textbf{\textcolor[rgb]{0.3,0.3,0.3}{0.583 (0.39)}} & 3.411 (1.896) & \textbf{\textcolor[rgb]{0.3,0.3,0.3}{2.639 (1.054)}} & 1.437 (0.617) \\ 
$\nu_{ \mbox{\tiny{DPI}};5}$  & 0.217 (0.359) & \textbf{\textcolor[rgb]{0.3,0.3,0.3}{0.636 (0.478)}} & 1.49 (1.219) & 0.737 (0.558) & \textbf{3.138 (1.783)} & 2.874 (1.398) & 1.268 (0.684) \\ 
$\nu_{ \mbox{\tiny{STE}};1}$ & 0.383 (0.377) & 0.708 (0.5) & 1.512 (1.13) & 0.678 (0.486) & \textbf{\textcolor[rgb]{0.3,0.3,0.3}{3.188 (2.162)}} & \textbf{2.614 (1.122)} & \textbf{1.172 (0.606)} \\  
$\nu_{ \mbox{\tiny{RT}}}$ & \textbf{0.017 (0.055)} & 0.646 (0.511) & \textbf{1.26 (0.89)} & \textbf{0.536 (0.364)} & 8.587 (2.629) & 3.173 (0.962) & 10.554 (0.368) \\ 
$\nu_{ \mbox{\tiny{MvM}}}$ & 0.643 (1.007) & 1.077 (1.126) & 1.998 (1.789) & 1.118 (1.157) & 3.26 (1.883) & 3.27 (1.738) & 1.449 (1.051) \\ 
$\nu_{ \mbox{\tiny{LCV}}}$ & 0.334 (0.698) & 0.73 (0.589) & 1.51 (1.215) & 0.731 (0.632) & 6.827 (2.769) & 2.859 (1.241) & \textbf{\textcolor[rgb]{0.3,0.3,0.3}{1.259 (0.69)}} \\ 
    \hline
Model & M8 & M9 & M10 & M11 & M12 & M13 & M14 \\ 
$n=50$  & \includegraphics[width=20mm]{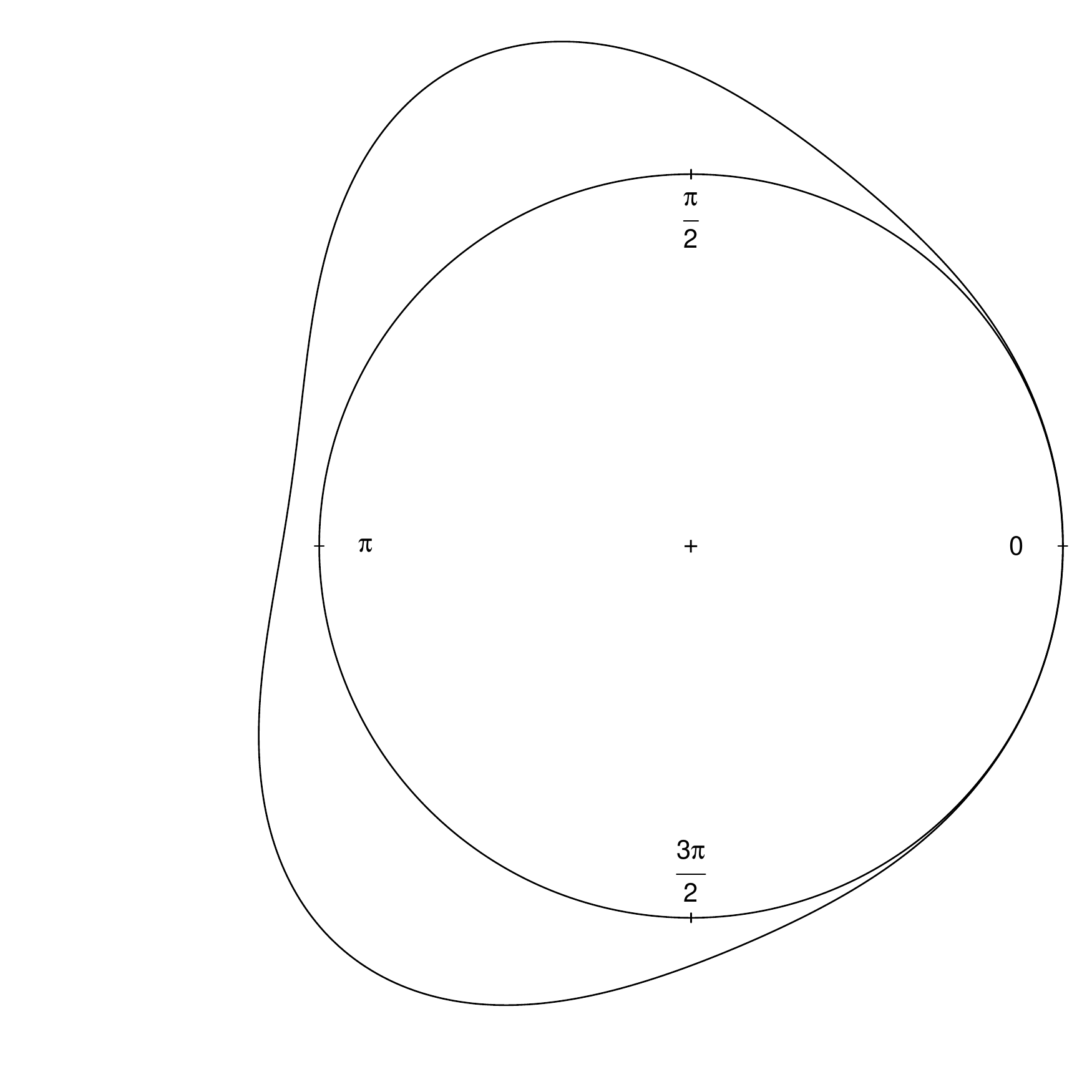} & \includegraphics[width=20mm]{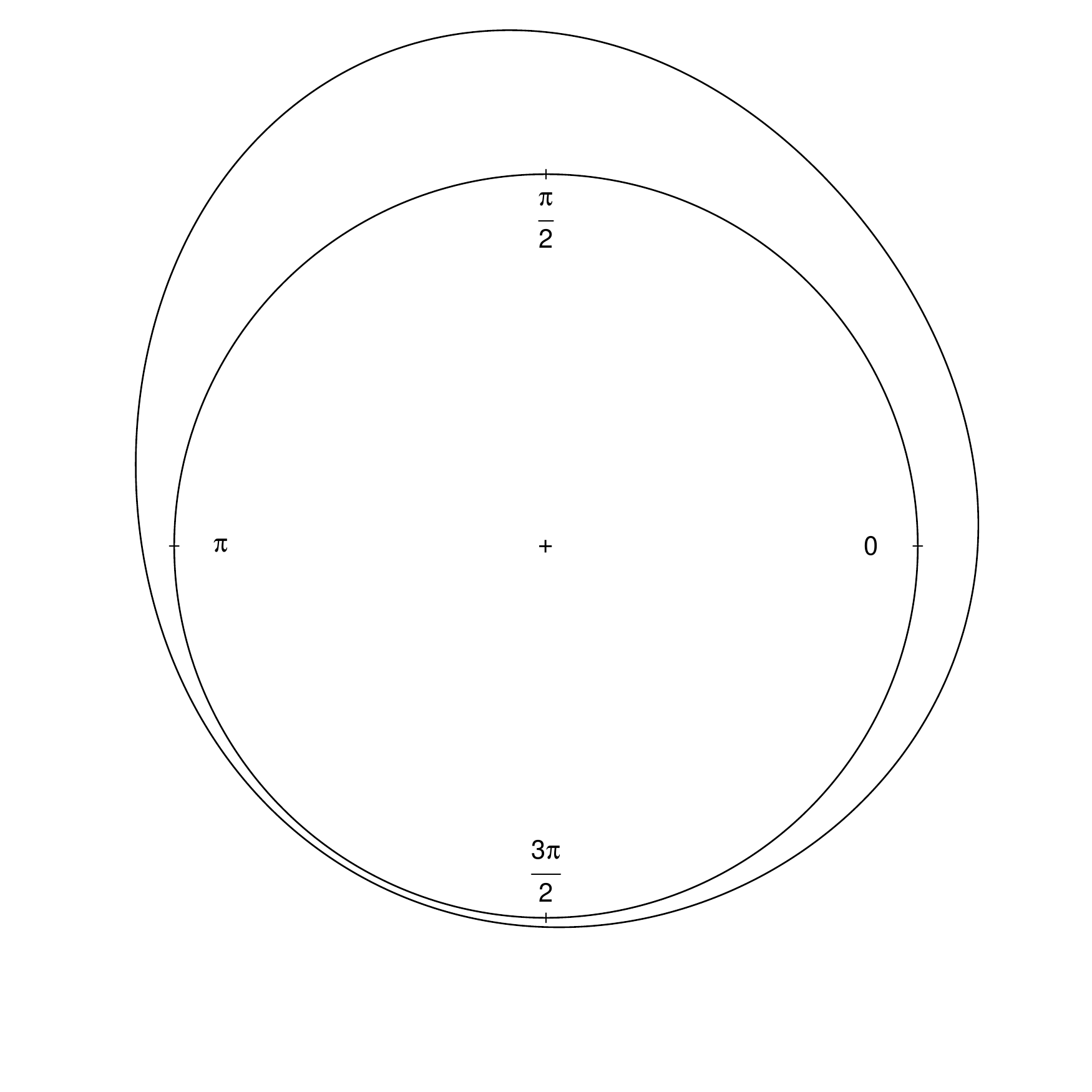}& \includegraphics[width=20mm]{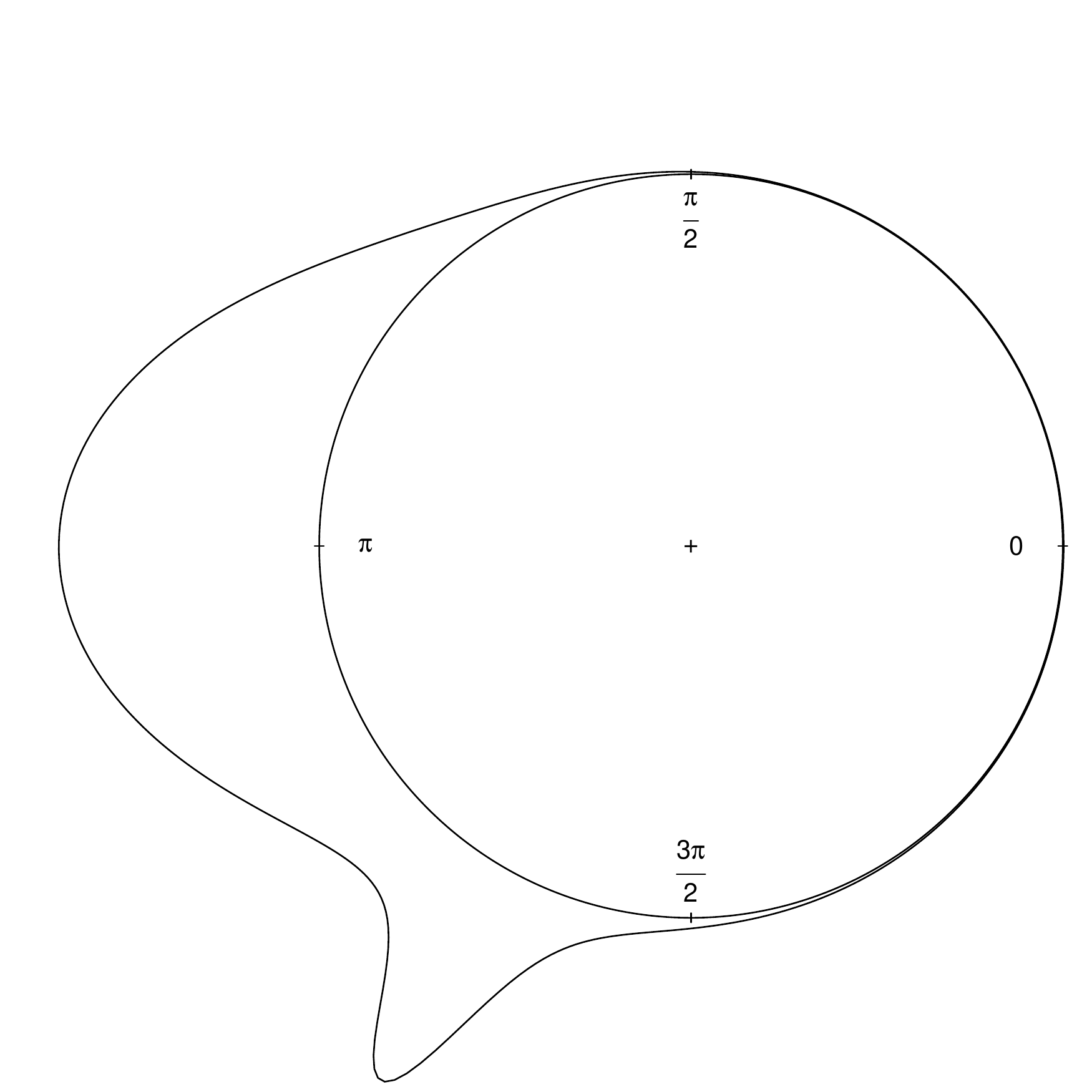}& \includegraphics[width=20mm]{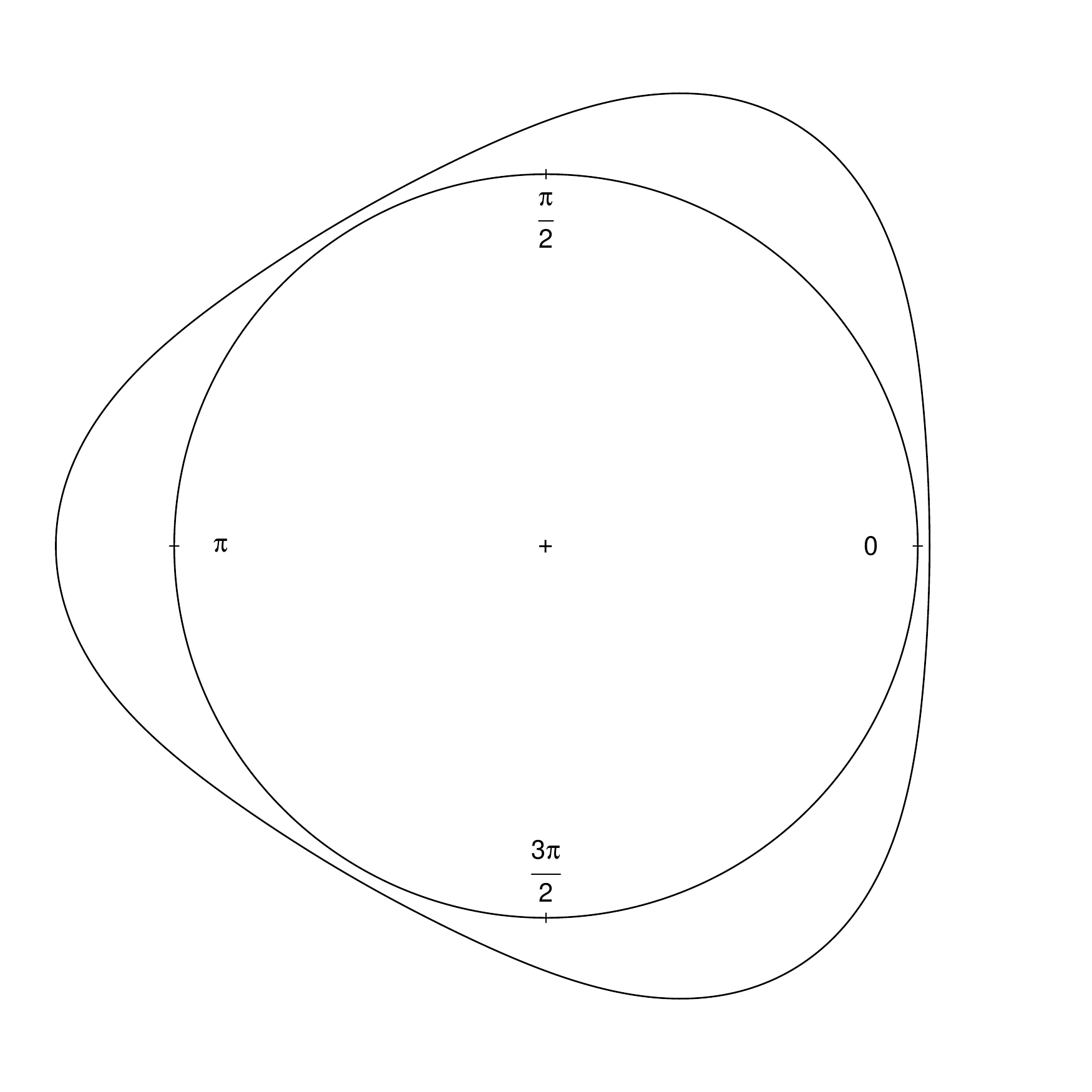}& \includegraphics[width=20mm]{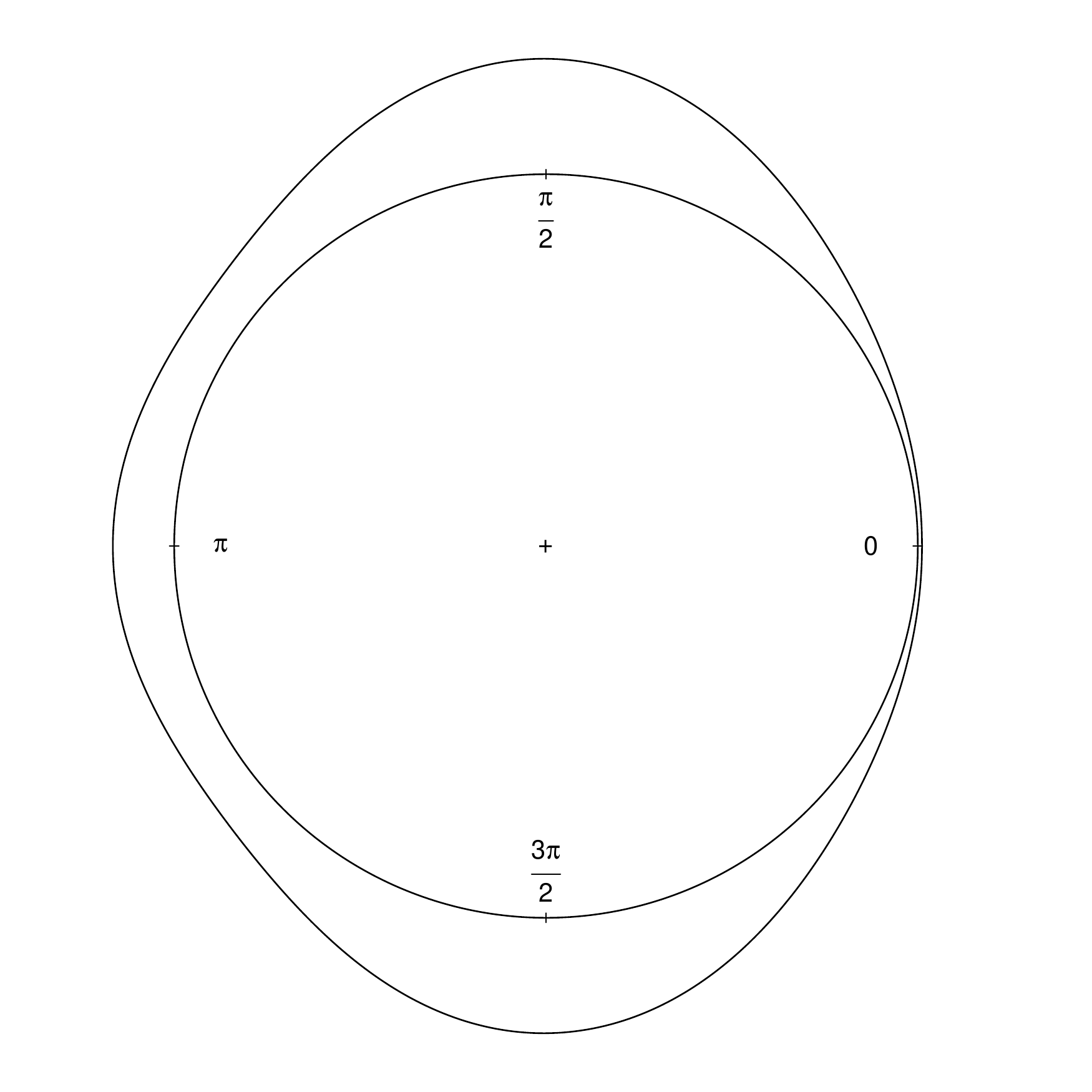}& \includegraphics[width=20mm]{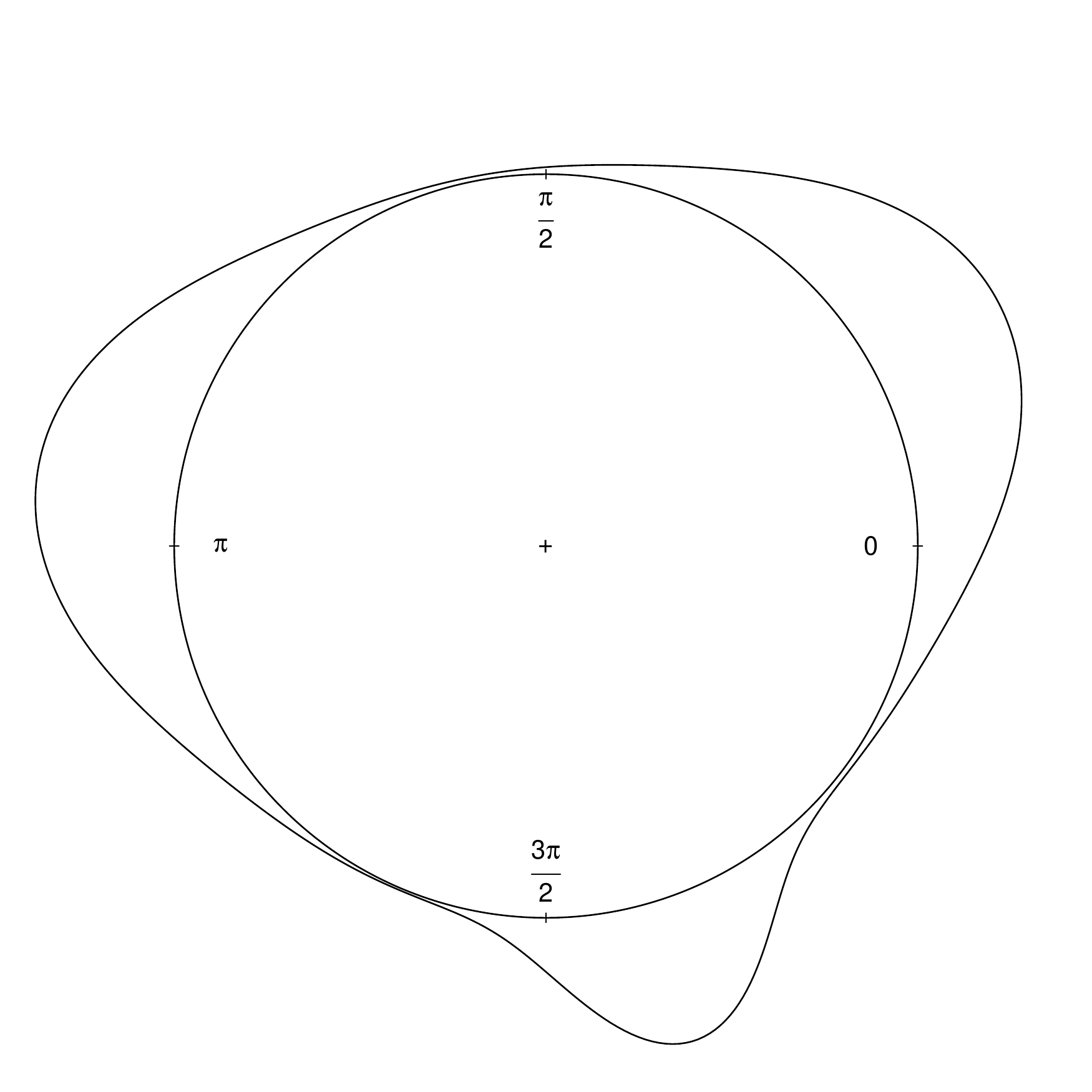}& \includegraphics[width=20mm]{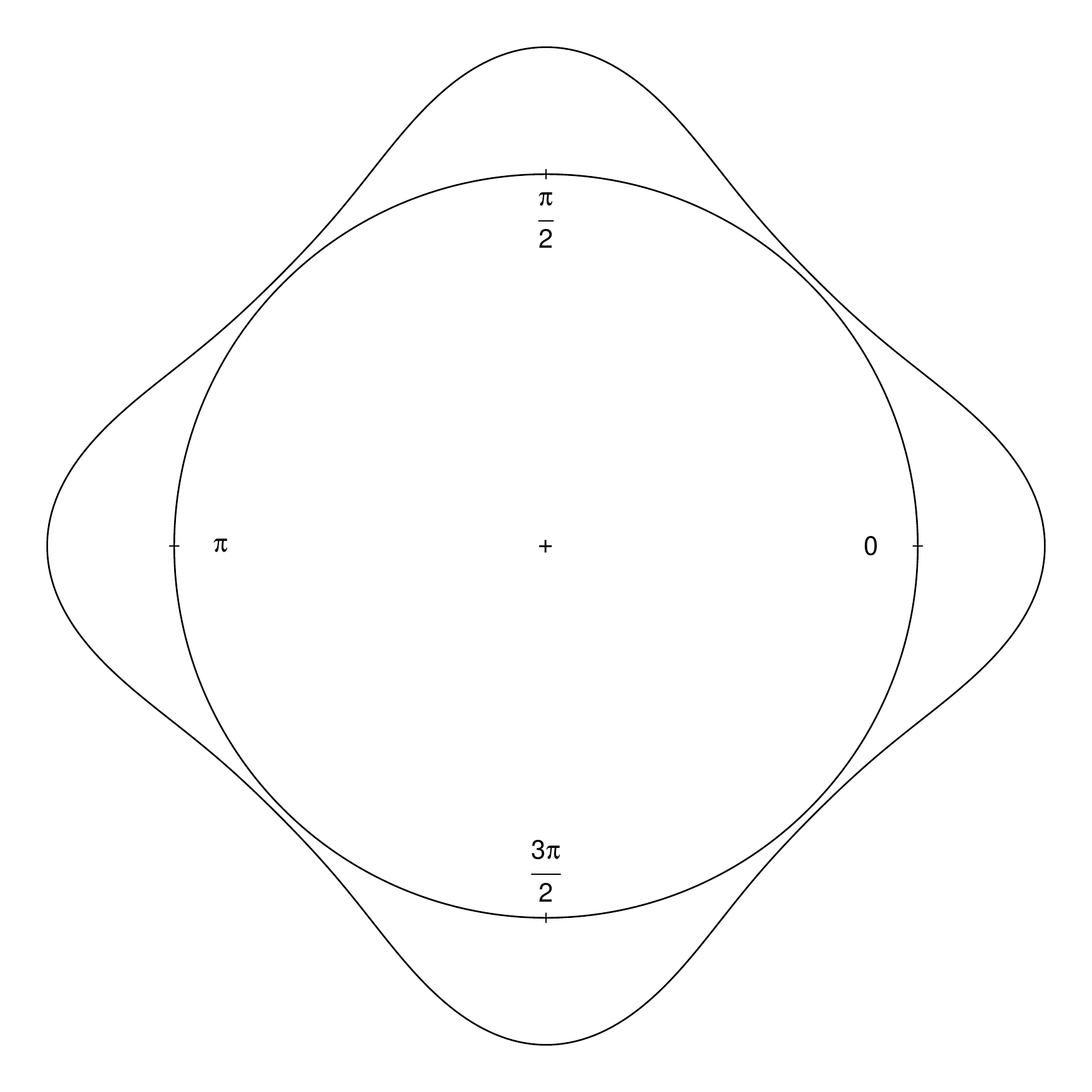} \\
$\nu_{ \mbox{\tiny{GS}}}$ & 1.981 (1.173) & 0.967 (0.688) & 3.141 (1.303) & 2.213 (1.014) & 1.563 (0.796) & 2.742 (1.217) & 3.035 (1.287) \\ 
$\nu_{ \mbox{\tiny{DPI}};1}$  & \textbf{2.289 (1.284)} & \textbf{1.259 (0.86)} & \textbf{3.64 (1.477)} & 3.99 (1.069) & \textbf{\textcolor[rgb]{0.3,0.3,0.3}{1.925 (0.803)}} & 5.229 (1.354) & 7.44 (0.767) \\ 
$\nu_{ \mbox{\tiny{DPI}};5}$  & 2.498 (1.685) & 1.449 (1.152) & 4.195 (2.203) & 2.801 (1.429) & 2.13 (1.238) & 3.389 (1.689) & 3.908 (1.832) \\ 
$\nu_{ \mbox{\tiny{STE}};1}$ & \textbf{\textcolor[rgb]{0.3,0.3,0.3}{2.29 (1.436)}} & 1.487 (1.102) & \textbf{\textcolor[rgb]{0.3,0.3,0.3}{3.799 (1.912)}} & \textbf{2.418 (1.095)} & \textbf{1.822 (0.926)} & \textbf{2.955 (1.259)} & \textbf{3.244 (1.344)} \\ 
$\nu_{ \mbox{\tiny{RT}}}$ & 4.921 (0.962) & \textbf{\textcolor[rgb]{0.3,0.3,0.3}{1.329 (0.956)}} & 3.811 (1.29) & 6.508 (0.075) & 4.451 (0.521) & 10.865 (0.142) & 8.226 (0.152) \\ 
$\nu_{ \mbox{\tiny{MvM}}}$ & 3.25 (3.033) & 2.649 (2.839) & 4.925 (3.217) & 3.37 (2.357) & 3.211 (2.756) & 4.321 (2.784) & 4.42 (2.501) \\ 
$\nu_{ \mbox{\tiny{LCV}}}$ & 2.454 (1.481) & 1.351 (1.045) & 4.152 (1.58) & \textbf{\textcolor[rgb]{0.3,0.3,0.3}{2.589 (1.245)}} & 1.991 (1.102) & \textbf{\textcolor[rgb]{0.3,0.3,0.3}{3.088 (1.349)}} & \textbf{\textcolor[rgb]{0.3,0.3,0.3}{3.382 (1.394)}} \\ 
   \hdashline
$n=100$ & &&&&&&\\ 
$\nu_{ \mbox{\tiny{GS}}}$ & 1.237 (0.698) & 0.6 (0.415) & 2.237 (0.844) & 1.313 (0.571) & 0.968 (0.485) & 1.7 (0.696) & 1.803 (0.702) \\ 
$\nu_{ \mbox{\tiny{DPI}};1}$  & \textbf{\textcolor[rgb]{0.3,0.3,0.3}{1.393 (0.745)}} & \textbf{0.739 (0.458)} & \textbf{\textcolor[rgb]{0.3,0.3,0.3}{2.572 (0.87)}} & 2.638 (0.905) & 1.191 (0.519) & 3.458 (1.045) & 6.393 (1.207) \\ 
$\nu_{ \mbox{\tiny{DPI}};5}$  & 1.466 (0.838) & 0.818 (0.544) & 2.726 (1.058) & 1.497 (0.654) & 1.188 (0.594) & 1.941 (0.824) & 2.061 (0.817) \\ 
$\nu_{ \mbox{\tiny{STE}};1}$ & \textbf{1.384 (0.752)} & 0.82 (0.526) & \textbf{2.552 (0.927)} & \textbf{1.405 (0.6)} & \textbf{1.082 (0.533)} & \textbf{1.824 (0.708)} & \textbf{1.904 (0.731)} \\  
$\nu_{ \mbox{\tiny{RT}}}$ & 3.727 (0.678) & \textbf{\textcolor[rgb]{0.3,0.3,0.3}{0.781 (0.51)}} & 2.895 (0.792) & 6.485 (0.014) & 4.265 (0.523) & 10.874 (0.125) & 8.182 (0.043) \\ 
$\nu_{ \mbox{\tiny{MvM}}}$ & 1.671 (1.207) & 1.102 (1.039) & 3.093 (1.433) & 1.719 (1.022) & 1.454 (0.999) & 2.131 (1.052) & 2.189 (1.01) \\ 
$\nu_{ \mbox{\tiny{LCV}}}$ & 1.49 (0.821) & 0.809 (0.536) & 3.08 (1.038) & \textbf{\textcolor[rgb]{0.3,0.3,0.3}{1.492 (0.664)}} & \textbf{\textcolor[rgb]{0.3,0.3,0.3}{1.171 (0.626)}} & \textbf{\textcolor[rgb]{0.3,0.3,0.3}{1.856 (0.732)}} & \textbf{\textcolor[rgb]{0.3,0.3,0.3}{1.969 (0.78)}} \\ 
   \hline
\end{tabular}}
\caption{Average ISE ($\times 100$) and standard deviations ($\times 100$, in parentheses) computed from 1000 samples of sample size $n=50$ (first block) and $n=100$ (second block) of Models M1--M14. In bold, smallest (dark) and second smallest (grey) average ISE. Smoothing parameters: concentration minimizing the ISE for each sample ($\nu_{ \mbox{\tiny{GS}}}$), proposed direct plug-in rule with $M_{\max}=1$ $(\nu_{ \mbox{\tiny{DPI}};1})$, direct plug-in rule with $M_{\max}=5$ $(\nu_{ \mbox{\tiny{DPI}};5})$, solve-the-equation plug-in selector with $M_{\max}=1$ $(\nu_{ \mbox{\tiny{STE}};1})$, rule of thumb of \cite{Taylor2008} $(\nu_{ \mbox{\tiny{RT}}})$, plug-in rule of \cite{Oliveira2012} $(\nu_{ \mbox{\tiny{MvM}}})$, and likelihood cross-validation of \cite{Hall1987} $(\nu_{ \mbox{\tiny{LCV}}})$.}
\label{table_results1}
\end{table}

\begin{table}[ht]
\centering
\scalebox{0.85}{
\begin{tabular}{lllllll}
  \hline
Model & M15 & M16 & M17 & M18 & M19 & M20   \\ 
$n=50$  & \includegraphics[width=20mm]{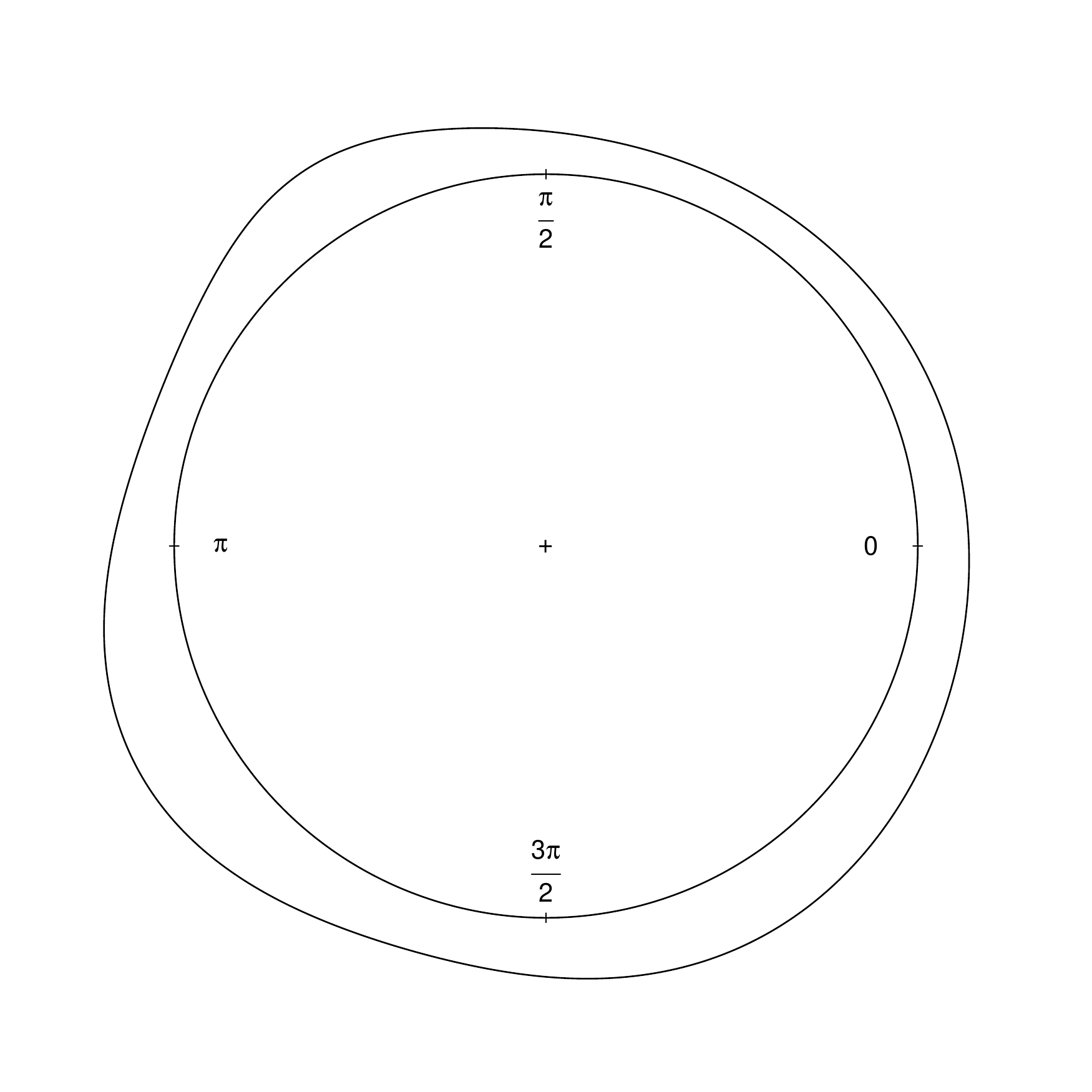} & \includegraphics[width=20mm]{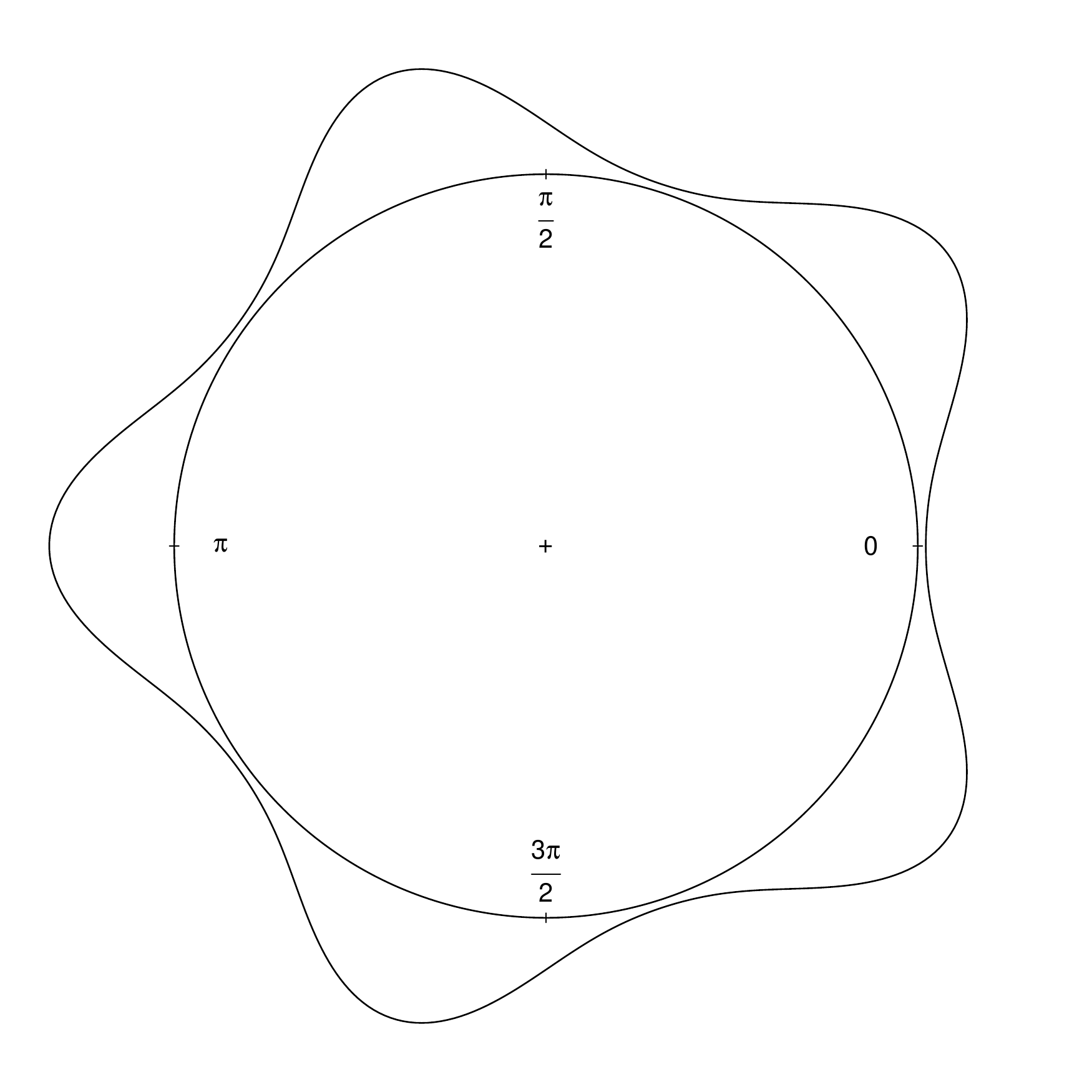}& \includegraphics[width=20mm]{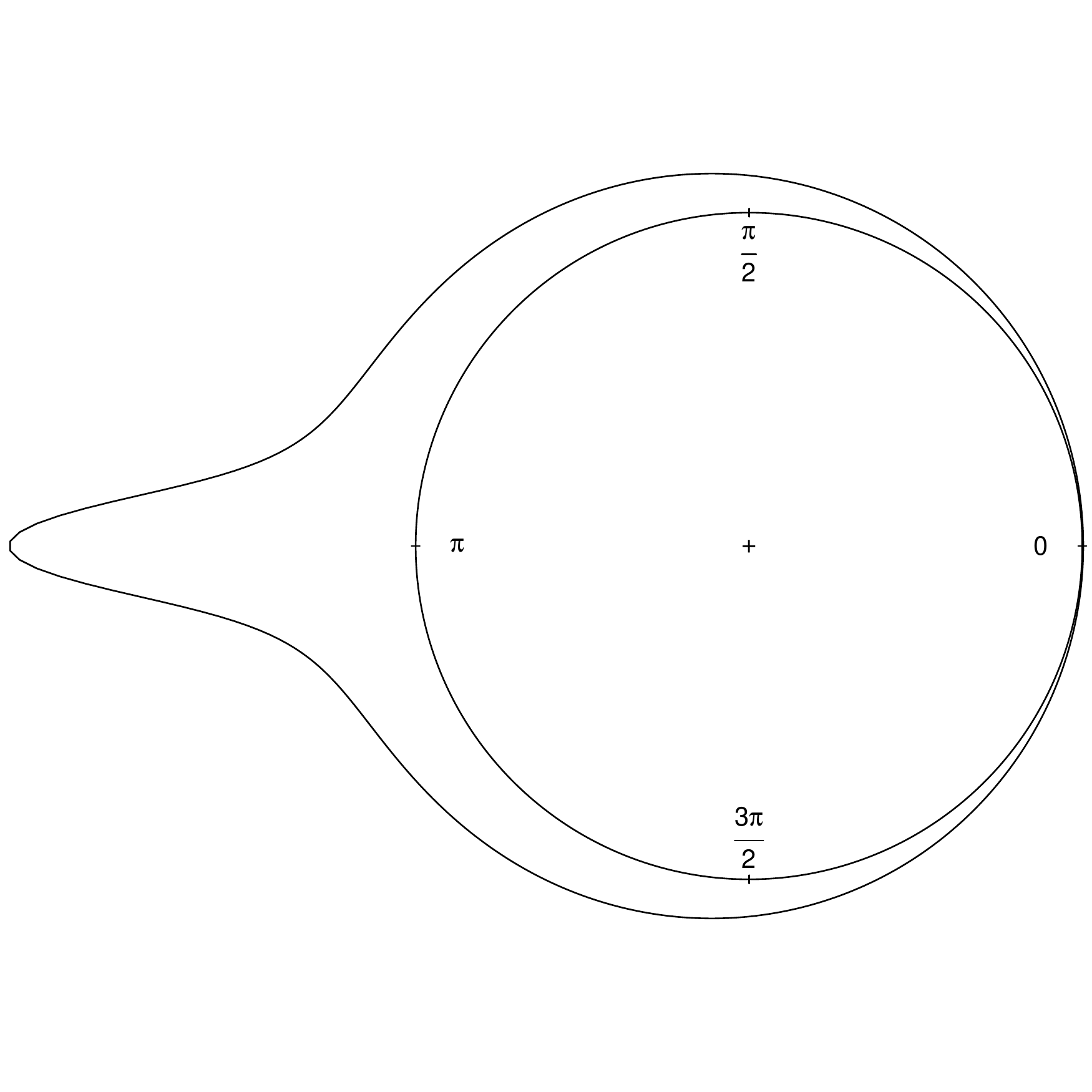}& \includegraphics[width=20mm]{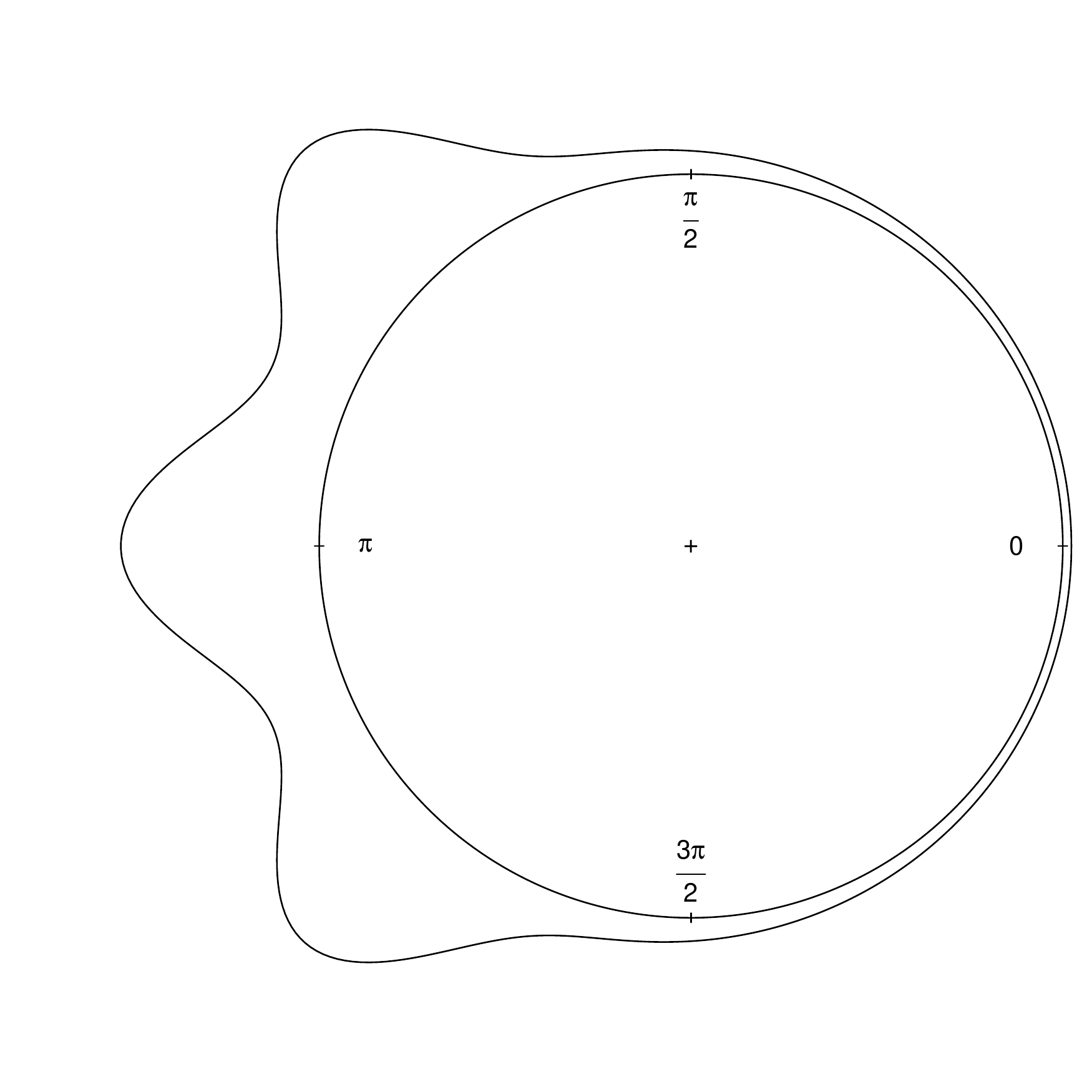}& \includegraphics[width=20mm]{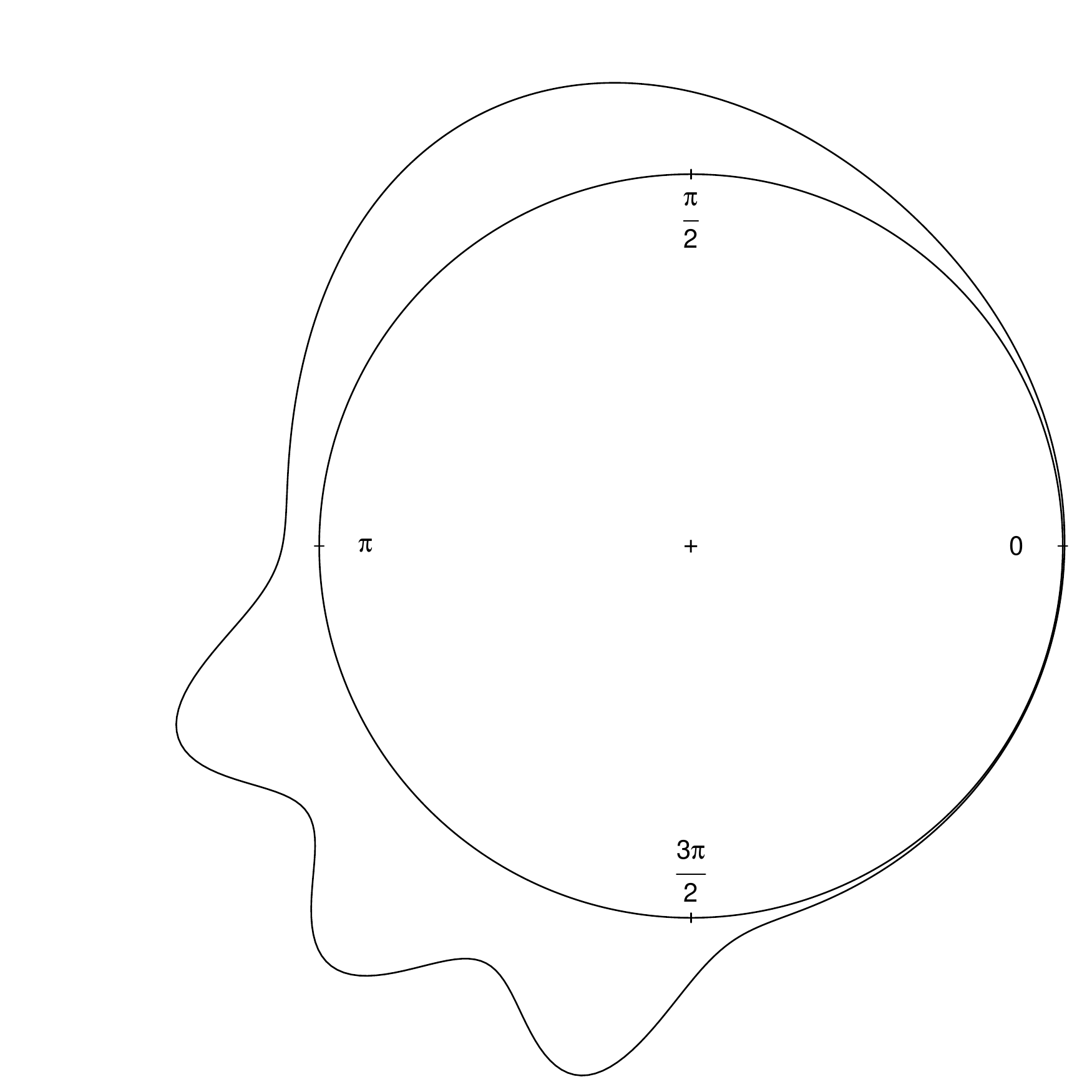}& \includegraphics[width=20mm]{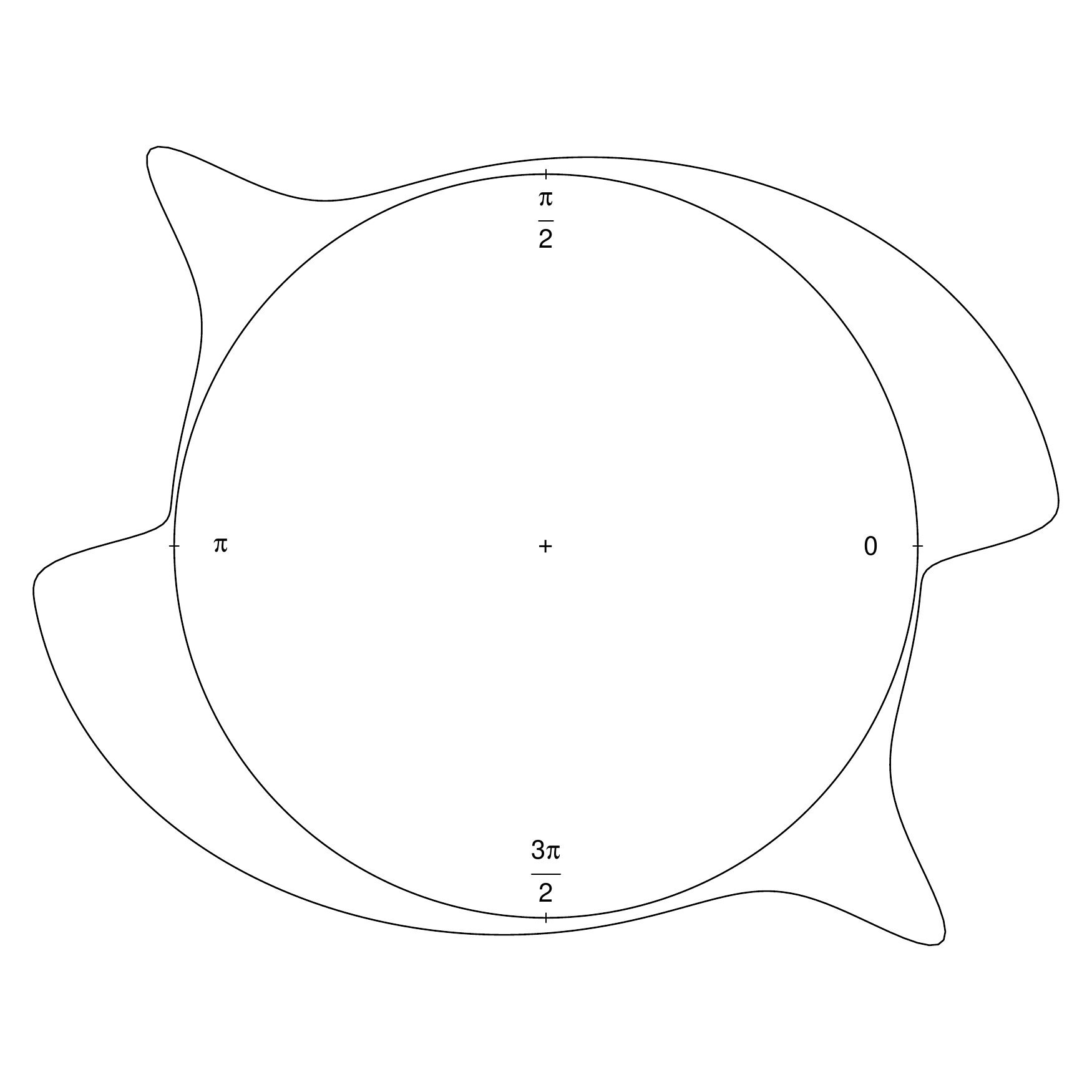}  \\
$\nu_{ \mbox{\tiny{GS}}}$ & 0.743 (0.201) & 3.644 (1.313) & 5.207 (1.906) & 3.211 (1.067) & 3.03 (1.001) & 4.965 (1.385) \\  
$\nu_{ \mbox{\tiny{DPI}};1}$  & \textbf{\textcolor[rgb]{0.3,0.3,0.3}{1.055 (0.519)}} & 8.128 (0.345) & 6.886 (2.07) & \textbf{\textcolor[rgb]{0.3,0.3,0.3}{3.949 (0.976)}} & \textbf{3.394 (1.027)} & 9.17 (0.836) \\ 
$\nu_{ \mbox{\tiny{DPI}};5}$  & 1.256 (0.968) & 4.942 (2.071) & \textbf{\textcolor[rgb]{0.3,0.3,0.3}{6.821 (2.155)}} & 4.067 (1.204) & 3.589 (1.579) & 6.113 (2.155) \\ 
$\nu_{ \mbox{\tiny{STE}};1}$ & 1.404 (0.829) & \textbf{\textcolor[rgb]{0.3,0.3,0.3}{4.017 (1.411)}} & \textbf{6.162 (2.224)} & \textbf{3.815 (1.306)} & \textbf{\textcolor[rgb]{0.3,0.3,0.3}{3.41 (1.227)}} & \textbf{\textcolor[rgb]{0.3,0.3,0.3}{5.914 (1.364)}} \\
$\nu_{ \mbox{\tiny{RT}}}$ & \textbf{0.903 (0.192)} & 7.879 (0.17) & 8.98 (1.667) & 4.212 (0.905) & 4.768 (1.004) & 10.966 (0.11) \\ 
$\nu_{ \mbox{\tiny{MvM}}}$ & 2.781 (2.596) & 5.12 (2.396) & 7.196 (2.689) & 5.089 (2.562) & 4.737 (2.862) & 6.811 (2.565) \\ 
$\nu_{ \mbox{\tiny{LCV}}}$ & 1.364 (0.98) & \textbf{3.973 (1.425)} & 7.897 (2.241) & 4.11 (0.986) & 3.578 (1.268) & \textbf{5.472 (1.529)} \\ 
    \hdashline
$n=100$ & &&&&&\\ 
$\nu_{ \mbox{\tiny{GS}}}$ & 0.63 (0.193) & 2.152 (0.775) & 3.437 (1.259) & 2.159 (0.8) & 2.188 (0.636) & 3.26 (0.888) \\  
$\nu_{ \mbox{\tiny{DPI}};1}$  & \textbf{0.768 (0.246)} & 7.865 (0.14) & 5.196 (1.59) & 3.014 (0.64) & 2.569 (0.581) & 8.133 (0.815) \\ 
$\nu_{ \mbox{\tiny{DPI}};5}$  & 0.86 (0.403) & 2.611 (1.247) & 4.695 (1.767) & 2.982 (0.745) & \textbf{2.48 (0.749)} & \textbf{3.491 (0.977)} \\ 
$\nu_{ \mbox{\tiny{STE}};1}$ & \textbf{\textcolor[rgb]{0.3,0.3,0.3}{0.843 (0.374)}} & \textbf{2.324 (0.805)} & \textbf{4.36 (1.822)} & \textbf{2.59 (0.958)} & \textbf{\textcolor[rgb]{0.3,0.3,0.3}{2.483 (0.62)}} & 4.067 (0.893) \\  
$\nu_{ \mbox{\tiny{RT}}}$ & 0.85 (0.153) & 7.837 (0.045) & 7.93 (1.198) & 3.563 (0.493) & 3.875 (0.639) & 10.961 (0.053) \\ 
$\nu_{ \mbox{\tiny{MvM}}}$ & 1.379 (1.028) & 2.5 (1.008) & \textbf{\textcolor[rgb]{0.3,0.3,0.3}{4.389 (1.744)}} & 3.081 (1.336) & 2.937 (1.237) & 3.893 (1.238) \\ 
$\nu_{ \mbox{\tiny{LCV}}}$ & 0.971 (0.552) & \textbf{\textcolor[rgb]{0.3,0.3,0.3}{2.34 (0.849)}} & 5.695 (1.949) & \textbf{\textcolor[rgb]{0.3,0.3,0.3}{2.976 (0.83)}} & 2.519 (0.686) & \textbf{\textcolor[rgb]{0.3,0.3,0.3}{3.514 (0.899)}} \\ 
   \hline
\end{tabular}}
\caption{Average ISE ($\times 100$) and standard deviations ($\times 100$, in parentheses) computed from 1000 samples of sample size $n=50$ (first block) and $n=100$ (second block) of Models M15--M20. Smoothing parameters: concentration minimizing the ISE for each sample ($\nu_{ \mbox{\tiny{GS}}}$), proposed direct plug-in rule with $M_{\max}=1$ $(\nu_{ \mbox{\tiny{DPI}};1})$, direct plug-in rule with $M_{\max}=5$ $(\nu_{ \mbox{\tiny{DPI}};5})$, solve-the-equation plug-in selector with $M_{\max}=1$ $(\nu_{ \mbox{\tiny{STE}};1})$, rule of thumb of \cite{Taylor2008} $(\nu_{ \mbox{\tiny{RT}}})$, plug-in rule of \cite{Oliveira2012} $(\nu_{ \mbox{\tiny{MvM}}})$, and likelihood cross-validation of \cite{Hall1987} $(\nu_{ \mbox{\tiny{LCV}}})$.}
\label{table_results2}
\end{table}

As summary measures in Tables~\ref{table_results1} (for Models M1--M14) and~\ref{table_results2} (for Models M15--M20), for each rule for concentration selection, we show the average ISE and standard deviations, computed over 1000 replicates. The smallest average ISE values are highlighted in bold. On those tables, as a benchmark, we have also included the results with the ``gold standard'' smoothing parameter $\nu_{ \mbox{\tiny{GS}}}$, which is the value of $\nu$ minimizing the ISE$(\nu)$ for each generated sample. 

In Tables~\ref{table_results1} and~\ref{table_results2}, we can observe that the proposed plug-in rules are the ones providing the smallest or the second smallest average ISE. The obtained average ISE is also close to the benchmark, which is remarkable considering that no-extra information is given about the true density and sample sizes are ``small''. The ``smallest'' dispersion, measured through the standard deviation, is obtained either by the proposed plug-in rules ($\nu_{ \mbox{\tiny{DPI}};1}$ and $\nu_{ \mbox{\tiny{STE}};1}$) or by the rule of thumb $(\nu_{ \mbox{\tiny{RT}}})$.

In the following, we describe in more detail the behaviour of the two-stage plug-in rule in terms of the average ISE provided in Tables~\ref{table_results1} and~\ref{table_results2}. First, we can observe that the plug-in rule with a simple von Mises at stage 0 ($\nu_{ \mbox{\tiny{DPI}};1}$) provides, in general, the smallest average ISE, especially when the sample size is ``small'' ($n=50$). The first exception to this general pattern occurs for densities that are well-approximated by a von Mises (M1, M3, and M4), where, as expected, the rule of thumb $(\nu_{ \mbox{\tiny{RT}}})$ provides a slightly lower average ISE, being $\nu_{ \mbox{\tiny{DPI}};1}$ the second one providing the best results. The second and more important exception occurs for Models M7, M11, M13, M14, M16, and M20; whose associated densities are multimodal and (almost) $k$-fold rotational symmetric. As mentioned in Section~\ref{algorithm_two_stage}, the ``bad'' performance of $\nu_{ \mbox{\tiny{DPI}};1}$ is probably due to the uniform estimation of the density by the von Mises density at stage 0. As commented there, we can solve that issue of the two-stage plug-in rule by using the mixture of von Mises at stage 0 ($\nu_{ \mbox{\tiny{DPI}};5}$). For those densities, we can see that $\nu_{ \mbox{\tiny{DPI}};5}$ provides better results than $\nu_{ \mbox{\tiny{DPI}};1}$.

On the complex models, those densities that are multimodal (M7, M8, M10-M16, and M18--M20) or ``peaked'' (M5 and M17), in general, the smallest average ISE is obtained by the solve-the-equation plug-in smoothing selector $(\nu_{ \mbox{\tiny{STE}};1})$, especially when the sample size is ``moderate'' ($n=100$). Even on the strongly (reflectively) asymmetric model (M6), $\nu_{ \mbox{\tiny{STE}};1}$ provides the smallest average ISE ($n=100$). Thus, the only cases where $\nu_{ \mbox{\tiny{STE}};1}$ does not provide the smallest average ISE are the models well-approximated by a von Mises density (M1--M4 and M9). For the remaining models, if $\nu_{ \mbox{\tiny{STE}};1}$ is not the ``best'' smoothing selector (M5, M15, M19, and M20), it is the rule providing the second smallest average ISE after the direct plug-in rule ($\nu_{ \mbox{\tiny{DPI}}}$).

In summary, we have shown that the proposed plug-in rules provide competitive smoothing parameters for ``small/moderate'' sample sizes. We have also replicated the same simulation study for larger samples sizes ($n=250$, $n=500$, and $n=1000$). For the simple models (the ones well approximated by a von Mises density), $\nu_{ \mbox{\tiny{DPI}};1}$ and $\nu_{ \mbox{\tiny{RT}}}$ still provide the ``best'' results. In the remaining models, for $n=250$, we observe a similar pattern as that described for $n=100$, with $\nu_{ \mbox{\tiny{STE}};1}$ providing the ``best'' results most of the time. The relative performance (when compared with the other data-based smoothing parameters) of $\nu_{ \mbox{\tiny{DPI}};5}$ and $\nu_{ \mbox{\tiny{MvM}}}$ improves with the sample size. Regarding the solve-the-equation plug-in smoothing selector, as mentioned before, the ISE values with $M_{\max}=5$, for $n=50$ and $n=100$, were not shown since, in most of the cases, $M_{\max}=1$ provided better results. We observed a different behaviour, in the complex models, for larger sample sizes ($n=500$, and $n=1000$). In those cases, the solve-the-equation plug-in smoothing selector with $M_{\max}=5$ ($\nu_{ \mbox{\tiny{STE}};5}$) obtained lower ISE values than those derived with $M_{\max}=1$ ($\nu_{ \mbox{\tiny{STE}};1}$). As a result, for most of the non-simple models, $\nu_{ \mbox{\tiny{MvM}}}$ and $\nu_{ \mbox{\tiny{STE}};5}$ become the smoothing selectors providing the ``best'' results when $n=500$ and $n=1000$. In those scenarios, the selector $\nu_{ \mbox{\tiny{MvM}}}$ behaves better when data is generated by a mixture of von Mises, while $\nu_{ \mbox{\tiny{STE}};5}$ provides the best results in the mixtures with other (more complex) density models.  

As a summary of our findings in this simulation study, our recommendations are as follows.

\begin{itemize}
\item For small sample sizes (say $n=50$), employ the proposed direct plug-in rule with $M_{\max}=1$ ($\nu_{ \mbox{\tiny{DPI}};1}$). The only exception to this recommendation is if there could be evidence of the density function being multimodal and $k$-fold rotational symmetric. In that case, employ the solve-the-equation plug-in smoothing selector with $M_{\max}=1$ ($\nu_{ \mbox{\tiny{STE}};1}$).
\item For moderate sample sizes (say $n=100$ or $n=250$), employ the solve-the-equation plug-in smoothing selector with $M_{\max}=1$ ($\nu_{ \mbox{\tiny{STE}};1}$). The direct plug-in rule with $M_{\max}=1$ ($\nu_{ \mbox{\tiny{DPI}};1}$) is also recommendable if there is evidence of unimodality.
\item For large sample sizes (say $n=500$ or $n=1000$), employ the plug-in rule of \cite{Oliveira2012} ($\nu_{ \mbox{\tiny{MvM}}}$). An alternative would be the solve-the-equation plug-in smoothing parameter with $M_{\max}=5$ ($\nu_{ \mbox{\tiny{STE}};5}$). When compared with $\nu_{ \mbox{\tiny{MvM}}}$, the selector $\nu_{ \mbox{\tiny{STE}};5}$ should provide a slightly better estimation when the density presents a complex shape. 
\end{itemize}

\section{Real data application}\label{real_data}

In this section, we revisit the car accident data that can be found in \cite{ameijeiras2022}. This real dataset consists of the time of the day (at a resolution of one minute) at which the car crash happened in El Paso County (Texas, USA) in 2018. A total of 85 observations were recorded on the webpage of the National Highway Traffic Safety Administration of the United States. 

\begin{figure}[t]
  \centering
    \subfloat{
    \includegraphics[width=0.45\textwidth]{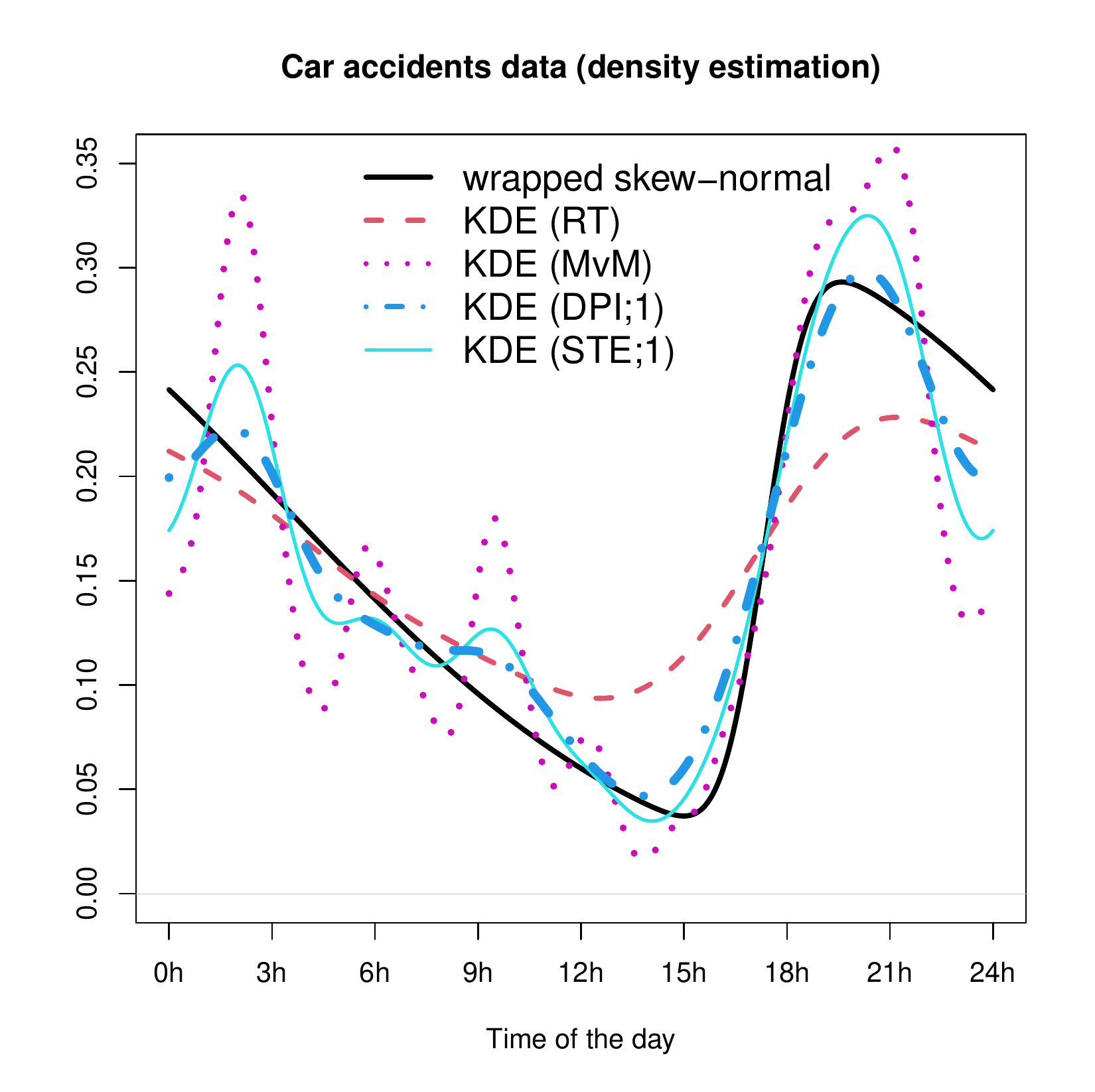}} \hspace{0.5cm}
    \subfloat{
    \includegraphics[width=0.45\textwidth]{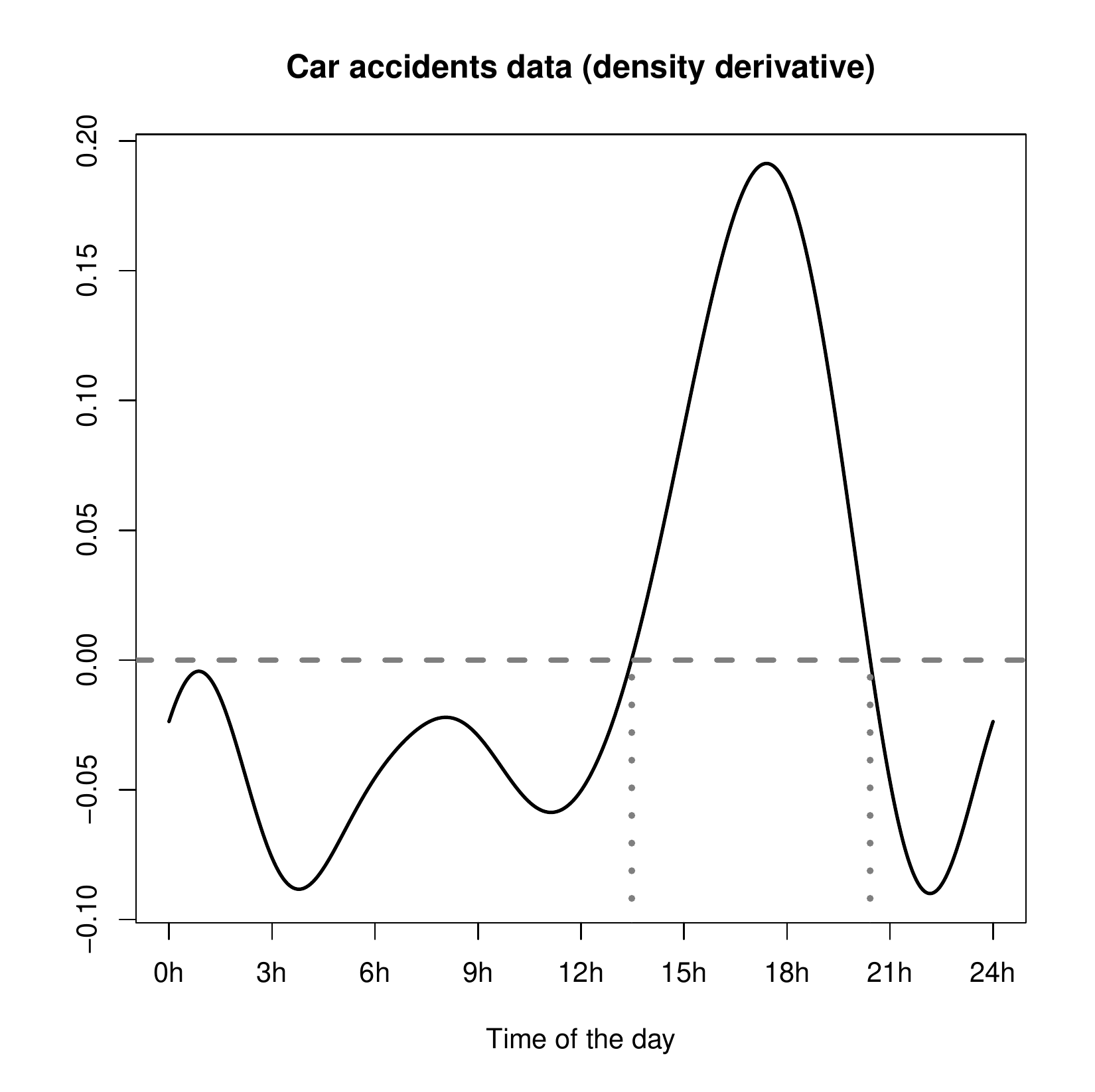}} 
  \caption{Car accident data. Left, density estimation: fitted wrapped skew normal density (thick solid line) and KDE with the von Mises kernel and different smoothing parameters. Smoothing parameters: rule of thumb of \cite{Taylor2008} (RT, dashed line), plug-in rule of \cite{Oliveira2012} (MvM, dotted line), proposed direct plug-in rule with $M_{\max}=1$ (DPI; 1; dot-dashed line), and proposed solve-the-equation plug-in rule with $M_{\max}=1$ (STE; 1; thin solid line). Right (solid line): kernel density derivative estimation, with a von Mises kernel and the proposed two-stage direct plug-in smoothing selector ($M_{\max}=1$). Right (dotted lines): estimated location of the modal and antimodal directions.}
\label{fig_car}
\end{figure}

In \cite{ameijeiras2022}, several features of the shape of the distribution are analysed with different inferential tools. The conclusion is that the density is unimodal and (reflectively) asymmetric. More specifically among several parametric models studied by \cite{ameijeiras2022}, the conclusion is that the ``best'' parametric fitting is achieved by the wrapped skew normal density of \cite{Pewsey2000}. A representation of the fitted wrapped skew normal density is provided in Figure~\ref{fig_car} (left, thick solid line).

This real dataset constitutes a good example where we would recommend employing the proposed direct plug-in smoothing selector, with $M_{\max}=1$, to estimate the density function as the sample size is ``small'' ($n=85$) and unimodality cannot be rejected for this sample. In Figure~\ref{fig_car} (left), we show the estimated density function employing this direct plug-in rule (DPI; 1), the proposed solve-the-equation plug-in selector (STE; 1), the rule of thumb of \cite{Taylor2008} (RT), and the plug-in rule of \cite{Oliveira2012} (MvM). We did not include the likelihood cross-validation of \cite{Hall1987} as its density estimation was close to the one obtained with the direct plug-in rule (DPI; 1). If we take the wrapped skew normal density as a reference, we can visually see that the closest kernel estimation is provided by the direct plug-in rule. This can be confirmed by computing the ISE~\eqref{ISE_nu}, replacing $f$ with the fitted wrapped skew normal density. In Table~\ref{tab_ise}, we can observe that the smallest ISE is obtained with the proposed direct plug-in rule.

\begin{table}[t]
\centering
\begin{tabular}{cccccc}
  \hline
Smoothing parameter & $\nu_{ \mbox{\tiny{RT}}}$ & $\nu_{ \mbox{\tiny{MvM}}}$ & $\nu_{ \mbox{\tiny{LCV}}}$ & $\nu_{ \mbox{\tiny{DPI}};1}$ & $\nu_{ \mbox{\tiny{STE}};1}$  \\ 
  \hline
ISE($\nu$)  ($\times 100$) & 1.134 & 1.765 & 0.325 & 0.265 & 0.509 \\ 
   \hline
\end{tabular}
\caption{ISE ($\times 100$) computed from the car accident data, taking the fitted wrapped skew normal density as the reference density $f$ in~\eqref{ISE_nu}. Smoothing parameters: rule of thumb of \cite{Taylor2008} $(\nu_{ \mbox{\tiny{RT}}})$, plug-in rule of \cite{Oliveira2012} $(\nu_{ \mbox{\tiny{MvM}}})$, likelihood cross-validation of \cite{Hall1987} $(\nu_{ \mbox{\tiny{LCV}}})$, proposed direct plug-in rule with $M_{\max}=1$ $(\nu_{ \mbox{\tiny{DPI}};1})$, and proposed solve-the-equation plug-in selector with $M_{\max}=1$ $(\nu_{ \mbox{\tiny{STE}};1})$.}
\label{tab_ise}
\end{table}
     
Finally, another application of the results derived in this paper can be found when the interest is in the density derivative estimate. As before, we will focus on the results obtained with the direct plug-in rule, with $M_{\max}=1$. In that case, the smoothing parameter for the kernel estimator can be derived with Algorithm~\ref{algorithm_proposed} (with $r=1$). The obtained estimator, with the von Mises kernel, for the car accident data is plotted in Figure~\ref{fig_car} (right). There, one can see that the derivative estimation is positive between 13:29 and 20:25 and it is negative during the remaining times of the day. Thus, if one wants to answer the question of when the peak of car accidents is produced, we see that, according to this estimator, one can find a modal direction at 20:25. The antimodal (valley) direction is achieved at 13:29. This is also remarkable as these values are close to the nonparametric modal (20:18) and antimodal (13:35) estimators by \cite{Ameijeiras-Alonso2019b}.

\section{Concluding remarks}\label{concluding}

The main contributions of this paper are the new plug-in smoothing parameters for circular kernel density (and its derivatives) estimation. In the past, some papers provided data-driven concentration parameters for estimating the circular density function, but still, a \cite{sheather1991} plug-in rule was missing in the circular literature. This paper fills that gap by providing all the theoretical results needed to derive an $l$-stage direct and solve-the-equation plug-in smoothing selectors for a general kernel satisfying some assumptions. The needed constants to obtain the optimal smoothing parameter are given for the ``most-popular'' circular kernels. Besides that, following the optimality criterion of \cite{muller1984}, this paper also discusses kernel choice. The conclusion is that the optimal kernel for circular density estimation is the wrapped Epanechnikov. This paper also includes a simulation study confirming that the proposed plug-in rules provide competitive smoothing parameters when compared with other proposals available in the statistical literature.

Although not done in this paper, the presented ideas could be extended to the multivariate toroidal setting, using the product kernel of \cite{DiMarzio2011}. In that case, the optimal smoothing parameter could be derived if the same kernel and concentration are employed in all the dimensions. Thus, in principle, one could obtain an explicit expression of the optimal smoothing parameter for the kernel estimators of the density derivative and the density functionals. Then, a similar scheme to that provided in Algorithms~\ref{algorithm_proposed} or~\ref{algorithm_proposed2} could be employed to obtain a two-stage or a solve-the-equation plug-in rule for the toroidal kernel density (and its derivatives) estimator.

\section{Availability}\label{software}

The proposed $l$-stage direct plug-in rule and the solve-the-equation selector have been added to the R library \texttt{NPCirc} \citep{Oliveira2014b}. Given a dataset \texttt{x}, the two-stage plug-in rule, described in Algorithm~\ref{algorithm_proposed} (with $M_{\max}=5$), for density estimation, can be obtained with the function \texttt{bw.AA(x, method="dpi")}. The solve-the-equation smoothing selector, described in Algorithm~\ref{algorithm_proposed2} (with $M_{\max}=1$), for density estimation is computed with \texttt{bw.AA(x)} or \texttt{bw.AA(x, method="ste")}. Thus, \texttt{bw.AA} is the circular equivalent of the function \texttt{bw.SJ} of the \texttt{stats} R package.

Some other extra possibilities are available on the function \texttt{bw.AA}. The kernel ($K$) and the derivative order ($r$ in \eqref{kernel_der}) can be selected, respectively, with the options \texttt{kernel} and \texttt{deriv.order}. If the practitioner prefers to use a different number of stages $l$, this can be selected with \texttt{nstage}. At stage 0, this function also allows for modifying the number of components $M$ (\texttt{M}) in the mixture~\eqref{mixture_von_Mises}, and selecting if a common concentration parameter is employed in all the components (\texttt{commonkappa}). Below, we provide the 3-stage direct plug-in rule for the first derivative of the density function. We employed the wrapped normal kernel and a mixture of 4 components, with different concentration parameters, at stage 0. We generated a sample \texttt{x} of 50 data from Model 18 from \cite{Oliveira2012}.

\begin{verbatim}
R> library("NPCirc")
R> x <- rcircmix(n=50,model=18)
R> bw.AA(x, deriv.order=1, method="dpi", nstage=3, kernel="wrappednormal",
+     M=4, commonkappa=FALSE) 
\end{verbatim}

For the direct plug-in rule, a slightly more accurate (and more computationally inefficient) concentration parameter could be obtained if the asymptotic approximations are avoided, see, e.g.,~\eqref{h_VM} and~\eqref{Q_VM}, for the von Mises kernel. This can be done if, at the end of every step of Algorithm~\ref{algorithm_proposed}, we compute an optimization routine searching the value of $\nu$ that minimizes the corresponding MISE/AMISE criterion~\eqref{amise_ker_der} or~\eqref{AMSE_psi}. In the optimization routine, we could employ as the initial value the concentration value achieved at each specific step (i.e., the values obtained by \eqref{optimal_h_psi} and \eqref{plugin_h}). This extra substep can be performed, in the \texttt{bw.AA} function, if we impose the argument \texttt{approximate=FALSE}. The simulation study of Section~\ref{simulation_study} was carried out also with this extra substep and similar results were obtained.  

A numerical routine is needed to obtain the value of the smoothing parameter, $\nu_{ \mbox{\tiny{STE}}}$, satisfying the Equality~\eqref{solve_h} in Step 4 of Algorithm~\ref{algorithm_proposed2}. In our simulations and in \texttt{bw.AA(x, method=`ste')}, the \texttt{uniroot} function of the \texttt{stats} R package was employed. The range over which the value of $h_{K; r; \mbox{\tiny{STE}}}$ is searched corresponds with the function arguments (\texttt{lower}, \texttt{upper}). The convergence tolerance, for searching the smoothing parameter with \texttt{uniroot}, can be modified with the argument \texttt{tol}.

In the function \texttt{bw.AA}, the values of the constants in \hyperref[cond7]{(E\ref{cond7})} and \hyperref[cond9]{(E\ref{cond9})} are only provided for the von Mises and the wrapped normal kernel. These values can be selected for other kernels with the arguments \texttt{Q1} (for $Q_{K;r,1}$) and \texttt{Q2} (for $Q_{K;r,2}$).

The kernel density derivative estimate for circular data was also added to the \texttt{NPCirc} package, inside the \texttt{kern.den.circ} function. By default it employs, as the smoothing parameter, the proposed solve-the-equation plug-in selector (with $M_{\max}=1$), described in Algorithm~\ref{algorithm_proposed2}. Alternatively, the concentration parameter (\texttt{bw}) can be a real value or a character related to the available data-based smoothing parameters (e.g., if \texttt{bw="dpi"}, the direct plug-in rule is employed). The derivative order can be selected with the argument \texttt{deriv.order}. This function generates an object of the class \texttt{density.circular} \citep[see the \texttt{circular} R package,][]{Agostinelli2017}. The remaining available options of \texttt{kern.den.circ} are, thus, similar to the ones in the function \texttt{density.circular}. A representation of the kernel density derivative estimator for the sample \texttt{x} could be obtained as follows.

\begin{verbatim}
R> fderhat <- kern.den.circ(x, deriv.order=1)
R> plot(fderhat, plot.type="line")
\end{verbatim}

The dataset containing the 85 car crashes that happened in El Paso County (Texas, USA) in 2018 is available on the website \url{https://github.com/jose-ameijeiras/Car-crashes-data}.

\appendix

\section{Asymptotic results for the kernel derivative estimator}\label{proofs_ker_der}

The main objective of this section is to compute the mean squared error (MSE) of the kernel derivative estimator in~\eqref{kernel_der}, i.e., $\mathbb{E}\left[\hat{f}^{(r)}_\nu(\theta) - {f}^{(r)}(\theta)\right]^2$. For doing so, firstly, we will consider the bias term and, secondly, the variance term. Regarding the expected value of $\hat{f}^{(r)}_\nu(\theta)$, we obtain the following result, using Taylor's theorem, Assumptions~\hyperref[cond1]{(A\ref{cond1})} over $f$, and \hyperref[cond2]{(A\ref{cond2})} over $K$.

\begin{align*}
\mathbb{E}\left[ \hat{f}^{(r)}_\nu(\theta) \right] &= \int_{-\pi}^{\pi}  K_\nu^{(r)}\left(\theta-\vartheta\right)  f(\vartheta) d \vartheta \nonumber \\
&= \int_{-\pi}^{\pi}  K_\nu\left(\varphi \right)  f^{(r)}(\theta+\varphi ) d \varphi  \nonumber \\
&= f^{(r)}(\theta)+ \frac{1}{2} \int_{-\pi}^{\pi} \varphi^2 K_\nu\left(\varphi \right)   d \varphi f^{(r+2)}(\theta) +O\left(\int_{-\pi}^{\pi} \varphi^4 K_\nu\left(\varphi \right)   d \varphi \right)
\end{align*}

Using the convergent Fourier series representation~\eqref{convergent_Fourier} and that $\int_{-\pi}^{\pi} \theta^2 \cos(j\theta) d \theta=4\pi (-1)^j/j^2$, for any $j$ integer, we obtain that $\int_{-\pi}^{\pi} \varphi^2 K_\nu\left(\varphi \right)   d \varphi= h_{K}(\nu)$. Thus, the bias of $\hat{f}^{(r)}_\nu(\theta)$ is equal to

\begin{equation*}
\mathbb{E}\left[ \hat{f}^{(r)}_\nu(\theta) \right] -f^{(r)}(\theta) = \frac{1}{2} h_{K}(\nu) f^{(r+2)}(\theta) +O\left(\int_{-\pi}^{\pi} \varphi^4 K_\nu\left(\varphi \right)   d \varphi \right).
\end{equation*} 

Now, if as $n\rightarrow\infty$, $h_{K}(\nu_n)=0$, i.e., if Assumption~\hyperref[cond3]{(A\ref{cond3})} is satisfied, $\hat{f}^{(r)}_\nu(\theta)$ is asymptotically unbiased and its asymptotic bias leading term is of order $O(h_{K}(\nu_n))$. If in addition, we assume Condition~\hyperref[cond4]{(E\ref{cond4})}, we obtain the following expression of the bias of $\hat{f}^{(r)}_\nu(\theta)$.

\begin{equation}\label{abias_ker_der}
\mbox{Bias}\left[\hat{f}^{(r)}_\nu(\theta)  \right] = \frac{1}{2} h_{K}(\nu_n) f^{(r+2)}(\theta) + o\left[h_{K}(\nu_n)\right].
\end{equation}

Regarding the variance term, using Assumptions~\hyperref[cond1]{(A\ref{cond1})}--\hyperref[cond3]{(A\ref{cond3})}, we obtain the following result.

\begin{align}\label{app1_3}
\mbox{Var}\left[ \hat{f}^{(r)}_\nu(\theta) \right] &= n^{-1} \int_{-\pi}^{\pi}  \left(K_\nu^{(r)}\left(\theta-\vartheta\right)\right)^2  f(\vartheta) d \vartheta - n^{-1} \left( \mathbb{E}\left[ \hat{f}^{(r)}_\nu(\theta) \right]  \right)^2 \nonumber \\
&= n^{-1}\int_{-\pi}^{\pi}  \left(K_\nu^{(r)}\left(\varphi \right) \right)^2  f(\theta+\varphi ) d \varphi - n^{-1} \left( f^{(r)}(\theta) + O(h_{K}(\nu))  \right)^2  \nonumber \\
&= n^{-1}\int_{-\pi}^{\pi}  \left(K_\nu^{(r)}\left(\varphi \right) \right)^2 d \varphi   f(\theta) +O\left(n^{-1}\int_{-\pi}^{\pi} \varphi^2 \left(K_\nu^{(r)}\left(\varphi \right) \right)^2 d \varphi \right) + O(n^{-1})
\end{align}

As before, employing the convergent Fourier series representation~\eqref{convergent_Fourier}, we would obtain that $R_{K;r,2}(\nu)=\int_{-\pi}^{\pi}  (K_\nu^{(r)}(\varphi ) )^2 d \varphi $. Now, by Assumption~\hyperref[cond3]{(A\ref{cond3})}, as $n\rightarrow\infty$, $n^{-1}=o[n^{-1}R_{K;r,2}(\nu_n)]$. Thus, the last term in \eqref{app1_3}, asymptotically vanishes. Then, the leading term of the asymptotic variance is $O(n^{-1}R_{K;r,2}(\nu_n))$. If the extra condition~\hyperref[cond8]{(E\ref{cond8})} for $K$ is also assumed, then the variance of $\hat{f}^{(r)}_\nu(\theta)$ is equal to the following quantity.

\begin{equation}\label{avar_ker_der}
\mbox{Var}\left[\hat{f}^{(r)}_\nu(\theta)  \right] =  n^{-1} R_{K;r,2}(\nu_n) f(\theta) + o\left[n^{-1}R_{K;r,2}(\nu_n)\right].
\end{equation}

Under the previous assumptions, combining~\eqref{abias_ker_der} and \eqref{avar_ker_der}, we derive the AMSE expression.

\begin{equation}\label{amse_ker_der}
\mbox{AMSE}\left[ \hat{f}^{(r)}_\nu (\theta)  \right] = \frac{1}{4} h_{K}^2(\nu_n)  \left(f^{(r+2)} (\theta)\right)^2 + \frac{1}{n} R_{K;r,2}(\nu_n) f(\theta).
\end{equation}

If $f^{(r+2)}$ is integrable, integrating the AMSE expression in~\eqref{amse_ker_der}, with respect to $\theta$, we derive the AMISE expression in~\eqref{amise_ker_der}. Now, the expression in~\eqref{optimal_h} of the optimal smoothing parameter with respect to the AMISE criterion would be obtained as follows. First, employing the Condition~\hyperref[cond5]{(E\ref{cond5})}, $R_{K;r,2}(\nu_n)=Q_{K;r,2} h_{n}^{-(2r+1)/2}$. Replacing that last value in~\eqref{amise_ker_der} and differentiating its expression with respect to $h_{n}$, if we set this derivative equal to zero, we obtain as the equation solution the value in~\eqref{optimal_h}.

\section{Asymptotic results for the kernel density functional}\label{proofs_functional}

The first objective of this section is to derive the MSE of $\hat \psi_{s; \rho}$, i.e., $\mathbb{E}\left[\hat \psi_{s; \rho} - \psi_{s}\right]^2$. For doing so, first consider the following reparametrization of the kernel estimator given in \eqref{psi_estimator},

\begin{equation}\label{app2_1}
\hat \psi_{s;\rho} = n^{-1} L_\rho^{(s)}\left( 0\right) + n^{-2} \overset {n} {\underset {i=1} \sum} \overset {n} {\underset {\substack{j=1 \\ j \neq i }} \sum} L_\rho^{(s)}\left(\Theta_i-\Theta_j\right).
\end{equation}

Since the first term does not depend on the data, we have only to study in detail the second term on the right-hand side of~\eqref{app2_1}. Using Taylor's theorem and Assumptions~\hyperref[cond1]{(A\ref{cond1})} over $f$ and \hyperref[cond2]{(A\ref{cond2})} over $L$ (when considering $s=r$), we obtain the following result.

\begin{align}\label{app2_2}
\mathbb{E}\left[ L_\rho^{(s)}\left(\Theta_1-\Theta_2\right) \right] &= \int_{-\pi}^{\pi} \int_{-\pi}^{\pi}  L_\rho^{(s)}\left(\theta-\vartheta\right) f(\theta) f(\vartheta) d\theta d \vartheta \nonumber \\
&= \int_{-\pi}^{\pi} \int_{-\pi}^{\pi}  L_\rho\left(\varphi \right) f(\varphi +\vartheta) f^{(s)}(\vartheta) d\varphi d \vartheta \nonumber \\
&= \psi_{s} + \frac{1}{2} h_{L} (\rho) \psi_{s+2} + O\left( \int_{-\pi}^{\pi} \theta^4 L_{\rho}(\theta) d\theta \right).
\end{align}

Now, combining~\eqref{app2_1} and~\eqref{app2_2}, under Assumption~\hyperref[cond3]{(A\ref{cond3})} (using $L$, instead of $K$; $\rho_n$, instead $\nu_n$), we obtain that the bias of $\hat \psi_{s; \rho}$ is equal to,

\begin{align*}
\mathbb{E}\left[\hat \psi_{s; \rho}\right]  - \psi_{s}  &=  n^{-1} L_\rho^{(s)}\left( 0\right) + (1-n^{-1}) \mathbb{E}\left[ L_\rho^{(s)}\left(\Theta_1-\Theta_2\right) \right] - \psi_{s}  \nonumber\\
&= n^{-1} \left( L_\rho^{(s)}\left( 0\right) + \psi_{s} \right) + \frac{1}{2} h_{L} (\rho) \psi_{s+2} + O\left( \int_{-\pi}^{\pi} \theta^4 L_{\rho}(\theta) d\theta \right)  + O\left(  n^{-1} h_{L} (\rho) \right).
\end{align*}

Now, using the convergent Fourier series representation~\eqref{convergent_Fourier}, it is easy to show that $L_\rho^{(s)}(0)= R_{L;s,1}(\rho)$. Therefore, under Assumptions~\hyperref[cond3]{(A\ref{cond3})}, \hyperref[cond6]{(A\ref{cond6})}, and~\hyperref[cond4]{(E\ref{cond4})}, we obtain that the bias is equal to the following quantity.

\begin{equation}\label{abias_psi}
\mbox{Bias}\left[ \hat \psi_{s; \rho} \right] =  n^{-1} R_{L;s,1}(\rho_n) + \frac{1}{2} h_{L} (\rho_n) \psi_{s+2} + o\left[ n^{-1} R_{L;s,1}(\rho_n) \right] + o\left[ h_{L} (\rho_n) \right].
\end{equation}

If Condition~\hyperref[cond4]{(E\ref{cond4})} is not fulfilled, then we can see that the asymptotic bias is of the same order, but we cannot derive a general explicit expression valid for any kernel $L$, $\mbox{Bias}\left[ \hat \psi_{s; \rho} \right] =  n^{-1} R_{L;s,1}(\rho_n) + O(h_{L} (\rho_n))$.

For the variance, consider again the reparametrization~\eqref{app2_1}. Now, since $L^{(s)}$ is symmetric for $s$ even, we have the following result \citep[see, e.g.,][Ch. 3]{WandJones1995}.

\begin{align}\label{var_psi}
\mbox{Var}\left[ \hat \psi_{s; \rho} \right] &=2 n^{-3}(n-1) \operatorname{Var}\left[L_\rho^{(s)} \left(\Theta_{1}-\Theta_{2}\right)\right] \nonumber \\
&+4 n^{-3}(n-1)(n-2) \operatorname{Cov}\left[L_\rho^{(s)}\left(\Theta_{1}-\Theta_{2}\right), L_\rho^{(s)}\left(\Theta_{2}-\Theta_{3}\right)\right].
\end{align}

Now, for computing the above value, first consider the following result, which is a consequence of Taylor's theorem, Assumption~\hyperref[cond1]{(A\ref{cond1})} over $f$, and \hyperref[cond2]{(A\ref{cond2})} over $L$ (when considering $s=r$).

\begin{align}\label{app2_3}
\mathbb{E}\left[ \left( L_\rho^{(s)}\left(\Theta_1-\Theta_2\right) \right)^2 \right] &= \int_{-\pi}^{\pi} \int_{-\pi}^{\pi}  \left(L_\rho^{(s)}\left(\theta-\vartheta\right) \right)^2 f(\theta) f(\vartheta) d\theta d \vartheta \nonumber \\
&= \int_{-\pi}^{\pi} \int_{-\pi}^{\pi}  \left( L_\rho\left(\varphi \right)\right)^2 f(\varphi +\vartheta) f^{(s)}(\vartheta) d\varphi d \vartheta \nonumber \\
&=  \psi_{0} R_{L;s,2}(\rho) + O\left( \int_{-\pi}^{\pi} \theta^2  L_\rho^2\left(\theta \right)d\theta \right).
\end{align}

Note that under Assumptions~\hyperref[cond1]{(A\ref{cond3})} and~\hyperref[cond8]{(E\ref{cond8})}, we obtain that, as $n\rightarrow\infty$, $\int_{-\pi}^{\pi} \theta^2  L_{\rho_n}^2\left(\theta \right)d\theta = o(R_{L;s,2}(\nu_n))$, and $n^{-1}R_{L;s,2}(\nu_n)=0$.

Similarly, the following result will be useful to derive the asymptotic results of the second part of the right-hand side of~\eqref{var_psi}.

\begin{align}\label{app2_4}
\mathbb{E}&\left[ L_\rho^{(s)}\left(\Theta_1-\Theta_2\right)  L_\rho^{(s)}\left(\Theta_2-\Theta_3\right)  \right] \nonumber\\
&\quad = \int_{-\pi}^{\pi} \int_{-\pi}^{\pi}  \int_{-\pi}^{\pi}  L_\rho^{(s)}\left(\theta-\vartheta\right) L_\rho^{(s)}\left(\vartheta-\varpi\right)  f(\theta) f(\vartheta) f(\varpi)d\theta d \vartheta d \varpi \nonumber \\
&\quad = \int_{-\pi}^{\pi} \int_{-\pi}^{\pi}  \int_{-\pi}^{\pi}  L_\rho\left(\varphi \right) L_\rho\left(\phi \right) f^{(s)}(\vartheta+\varphi )f(\vartheta)  f^{(s)}(\vartheta+\phi) d\varphi d \vartheta d\phi \nonumber \\
&\quad = \int_{-\pi}^{\pi} \left(f^{(s)}(\vartheta)\right)^2f(\vartheta)d \vartheta +O\left( h_{L} (\rho) \right).
\end{align}

The last term in~\eqref{app2_4} asymptotically vanishes using the Assumption~\hyperref[cond3]{(A\ref{cond3})} for the kernel $L$. From~\eqref{app2_2}, we obtain that $\mathbb{E}\left[ L_\rho^{(s)}\left(\Theta_1-\Theta_2\right) \right]= \psi_{s} + O(h_{L} (\rho))$. The variance of $\hat \psi_{s; \rho}$ can be obtained from~\eqref{var_psi}, by combining the results in \eqref{app2_2}, \eqref{app2_3}, and \eqref{app2_4}.

\begin{align}\label{avar_psi}
\mbox{Var}\left[ \hat \psi_{s; \rho} \right] &=2 n^{-2} \left( \psi_{0} R_{L;s,2}(\rho) + O\left( \int_{-\pi}^{\pi} \theta^2  L_{\rho}^2\left(\theta \right)d\theta \right) - \psi_s^2 \right) \nonumber \\
&+4 n^{-1} \left(\int_{-\pi}^{\pi} \left(f^{(s)}(\vartheta)\right)^2f(\vartheta)d \vartheta +O\left( h_{L} (\rho) \right)  - \psi_s^2\right)  \nonumber \\
\mbox{AVar}\left[ \hat \psi_{s; \rho} \right] &= 2 n^{-2} \psi_{0} R_{L;s,2}(\rho_n)+4 n^{-1} \left( \int_{-\pi}^{\pi} \left(f^{(s)}(\vartheta)\right)^2f(\vartheta)d \vartheta  - \psi_s^2 \right),
\end{align}
where the last equality is a consequence of Assumptions~\hyperref[cond1]{(A\ref{cond3})} and~\hyperref[cond8]{(E\ref{cond8})}.

Finally, the AMSE expression in~\eqref{AMSE_psi} is obtained by adding the square of the asymptotic bias~(see~\eqref{abias_psi}) and the asymptotic variance~\eqref{avar_psi}.

Note that under Conditions~\hyperref[cond4]{(E\ref{cond5})} and~\hyperref[cond7]{(E\ref{cond7})}, we obtain that $R_{L;s,2}(\rho_n)=o(R_{L;s,1}(\rho_n))$. Thus, the term depending on $\rho_n$ in the asymptotic variance~\eqref{avar_psi} is a small $o$ term of the squared asymptotic bias. The optimal value of $\mathfrak{h}_{L; n}$ in~\eqref{optimal_h_psi} can be obtained equating to zero the derivative of the square of the asymptotic bias (see~\eqref{abias_psi}), using the assumption about the sign in Condition~\hyperref[cond7]{(E\ref{cond7})}. 

When employing the smoothing parameter $\mathfrak{h}_{L; s; \mbox{\tiny{AMSE}}}$ the main term of the bias vanishes. Then, the minimal AMSE of $\hat \psi_{s; \rho}$ is of order

\begin{align*}
& O\left( \left( \int_{-\pi}^{\pi} \theta^4 L_{\rho_{L; s; \mbox{\tiny{AMSE}}}}(\theta) d\theta \right)^2 \right)   + O\left(  n^{-2} \mathfrak{h}_{L; s; \mbox{\tiny{AMSE}}}^2 (\rho) \right)+O(n^{-2} \mathfrak{h}_{L; s; \mbox{\tiny{AMSE}}}^{-(2r+1)/2}) + O(n^{-1}) \\
=& O\left( \left( \int_{-\pi}^{\pi} \theta^4 L_{\rho_{L; s; \mbox{\tiny{AMSE}}}}(\theta) d\theta \right)^2 \right)  + O(n^{-2(s+5)/(s+3)}) + O(n^{-5/(s+3)})+ O(n^{-1}) \\
=& O\left( \left( \int_{-\pi}^{\pi} \theta^4 L_{\rho_{L; s; \mbox{\tiny{AMSE}}}}(\theta) d\theta \right)^2 \right)  + O(n^{-5/(s+3)})+ O(n^{-1}). 
\end{align*}

Now, under Condition~\hyperref[cond9]{(E\ref{cond9})}, we obtain that, 

\begin{equation*}
\mbox{AMSE}\left[ \hat \psi_{s; \rho_{L; s; \mbox{\tiny{AMSE}}}} \right]=o(\mathfrak{h}_{L; s; \mbox{\tiny{AMSE}}}^{5/2})+ O(n^{-5/(s+3)})+ O(n^{-1})= O(n^{-5/(s+3)})+ O(n^{-1}).
\end{equation*}

The minimal AMSE orders in Corollary~\ref{cor_optimal_h_psi} follow. The explicit expression of the minimal AMSE can be obtained from the asymptotic variance expression in~\eqref{avar_psi}.

If Condition~\hyperref[cond9]{(E\ref{cond9})} is not satisfied, then the minimal AMSE is of the following order,

\begin{equation*}
\mbox{AMSE}\left[ \hat \psi_{s; \rho_{L; s; \mbox{\tiny{AMSE}}}} \right]=O\left( \left( \int_{-\pi}^{\pi} \theta^4 L_{\rho_{L; s; \mbox{\tiny{AMSE}}}}(\theta) d\theta \right)^2 \right)+ O(n^{-1})= o(n^{-4/(s+3)})+ O(n^{-1}).
\end{equation*}

\bibliographystyle{chicago}
\bibliography{DirectionalStats}   

\end{document}